%% file: main.tex
\documentclass[11pt]{article}

\usepackage[english]{babel}
\usepackage{graphicx, color, enumerate}
\usepackage{tkz-graph}
\usetikzlibrary{shapes.geometric} % necessary for \tikzstyle{VertexStyle}
\usetikzlibrary{fit} % Boxen um Knoten
\usepackage[onehalfspacing]{setspace}
\usepackage{adjustbox} % adjust table size to text width
\usepackage{amssymb,amsmath,amsthm,mathtools}
\usepackage{lscape} % Landscape of pages (Drehen um 90 Grad)
\usepackage{tocloft}
\usepackage{calrsfs} % Caligraphic math symbols
\usepackage{stackengine}% for wide hats in math environment
\usepackage{natbib}
\usepackage{longtable} % table with page breaks
\usepackage{float}
\usepackage[hypcap=false]{caption}
\usepackage{lscape}
\usepackage[hyphens,spaces,obeyspaces]{url}
\usepackage[splitrule,bottom]{footmisc}
\usepackage[textheight=722pt,textwidth=500pt,centering,headheight=50pt,headsep=12pt,footskip=12pt]{geometry}
\DeclareMathAlphabet{\pazocal}{OMS}{zplm}{m}{n} % define new caligraphic symbol
\makeatletter
\newcommand*{\rom}[1]{\expandafter\@slowromancap\romannumeral #1@}
\makeatother
\newcounter{z}
\def\z{\stepcounter{z}[[[[[\thez]]]]]}
\newcommand{\GG}[1]{} % a and b sorting of references with same year and author

\usepackage{bm}
\usepackage{bbm}
\usepackage[colorlinks,citecolor=purple!60!black,linkcolor=blue!60!black,urlcolor=purple!60!black]{hyperref}

\usepackage{caption} % captionof in Appendix

\usepackage{color, colortbl} % table colors
\definecolor{Gray}{gray}{0.95}
\usepackage{booktabs} % To thicken table lines
\newcolumntype{Y}{>{\centering\arraybackslash}X} % Center column in tabularx
\title{Estimating the Effect of Central Bank Independence on Inflation Using Longitudinal Targeted Maximum Likelihood Estimation\thanks{The views, opinions, findings, and conclusions or recommendations expressed in this paper are strictly those of the authors. They do not necessarily reflect the views of the Swiss National Bank (SNB)\@. The SNB takes no responsibility for any errors or omissions in, or for the correctness of, the information contained in this paper.}}
\author{Philipp F. M. Baumann\thanks{KOF Swiss Economic Institute, ETH Zurich. e-mail: baumann@kof.ethz.ch}
\and Michael Schomaker\thanks{Institute of Public Health, Medical Decision Making and Health Technology Assessment, UMIT - University for Health Sciences, Medical Informatics and Technology, Hall in Tirol, Austria and Centre for Infectious Disease Epidemiology \& Research, University of Cape Town, Cape Town, South Africa. e-mail: michael.schomaker@umit.at}
\and Enzo Rossi\thanks{Swiss National Bank and University of Zurich. e-mail: enzo.rossi@snb.ch}}

\date{\today}
\begin{document}

\maketitle
\pagenumbering{roman}
\thispagestyle{empty}

%\vspace{0.3cm}

\begin{abstract}
\noindent
The notion that an independent central bank reduces a country's inflation is a controversial hypothesis. To date, it has not been possible to satisfactorily answer this question because the complex macroeconomic structure that gives rise to the data has not been adequately incorporated into statistical analyses. We develop a causal model that summarizes the economic process of inflation. Based on this causal model and recent data, we discuss and identify the assumptions under which the effect of central bank independence on inflation can be identified and estimated. Given these and alternative assumptions, we estimate this effect using modern doubly robust effect estimators, i.e., longitudinal targeted maximum likelihood estimators. The estimation procedure incorporates machine learning algorithms and is tailored to address the challenges associated with complex longitudinal macroeconomic data. We do not find strong support for the hypothesis that having an independent central bank for a long period of time necessarily lowers inflation. Simulation studies evaluate the sensitivity of the proposed methods in complex settings when certain assumptions are violated and highlight the importance of working with appropriate learning algorithms for estimation.
\end{abstract}

%
%Journal classification:
%

\clearpage

\pagenumbering{arabic}
\section{Introduction}
The impact of the institutional design of central banks on real economic outcomes has received considerable attention over the past three decades. Whether central bank independence (CBI) can lower inflation and provide inflation stability in a country is a particularly controversial issue. It has been claimed that more than 9,000 works have been devoted to the investigation of the role of CBI in influencing economic outcomes \citep{vuletin2011replacing}. After the 2008-09 Global Financial Crisis, the debate on the optimal design of monetary policy authorities has become even more intense.

The statistical and economic literature is rich in studies that evaluate the relationship between CBI and inflation. A common approach is to treat countries as units in a linear regression model where inflation (the percentage change in the consumer price index, CPI) is the outcome and a binary CBI index and several economic and political variables are covariates. While many studies have found that an independent central bank may lower inflation \citep{grilli1991political, cukierman1992measuring, alesina1993central, klomp2010inflation, klomp2010central,  arnone2013dynamic}, other studies that have used a broader range of characteristics of a nation's economy have been unable to find such a relationship \citep{cargill1995statistical, fuhrer1997central,oatley1999central}. Moreover, there have been studies suggesting that the effect of CBI on inflation can only be seen during specific time periods \citep{klomp2010inflation} or only in developed countries \citep{klomp2010central,neyapti2012monetary,alpanda2014impact}.

Numerous articles have pointed out the weaknesses that come with simple cross-sectional regression approaches when evaluating the effect of CBI on inflation. First, the problem at hand is longitudinal in nature, and only an appropriate panel setup may be suitable to estimate the (long-term) effect of CBI on inflation. Second, the question of interest is essentially causal: i.e., what (average) inflation would we observe in 10 years' time, if -- from now on -- each country's monetary institution had an independent central bank compared to the situation in which the central bank were not independent? However, not all cross-sectional regression approaches embed their analyses in a holistic causal framework.

Some more recent work has attempted to overcome at least parts of these problems. For example, \citet{crowe2007evolution,crowe2008central} use a panel data setup with two time intervals, and \citet{klomp2010central} work with a random coefficient panel model to answer the question of interest in a longitudinal setup. Other authors, e.g., \citet{walsh2005optimal}, acknowledge not only that current CBI may cause future inflation but also that current inflation is possibly related to future CBI status. Several authors have thus tried to use instrumental variable approaches to estimate the effect of CBI on inflation within a causal framework, but have been unable to find strong instruments \citep{crowe2008central,jacome2008there}.

It is clear that evaluating the effect of CBI on inflation requires a longitudinal causal estimation approach. However, it has been shown repeatedly that standard regression approaches are typically not suitable to answer causal questions, particularly when the setup is longitudinal and when the time-dependent confounders of the outcome-intervention relationship are affected by previous intervention decisions (\citealp{greenland1998introduction, Daniel:2013}). There are at least three methods to evaluate the effect of longitudinal (multiple time-point) interventions on an outcome in such complex situations: 1) inverse probability of treatment weighted (IPTW) approaches (\citealp{Robins:2000}); 2) standardization with respect to the time-dependent confounders (i.e., g-formula-type approaches \citep{Robins:1986, Bang:2005}); and 3) doubly robust methods, such as targeted maximum-likelihood estimation (TMLE, \citealp{vanderLaan:2011}), which can be seen as a combination and generalization of the other two approaches.

Longitudinal targeted maximum likelihood estimation (LTMLE, \citealp{vanderLaan:2012}) is a doubly robust estimation technique that requires iteratively fitting models for the outcome and intervention mechanisms at each time point. With LTMLE, the causal quantity of interest (such as differences in counterfactual outcomes after intervening at multiple time points) is estimated consistently if either the iterated outcome regressions or the intervention mechanisms are estimated consistently. LTMLE, like other doubly robust methods, has an advantage over other approaches in that it can more readily incorporate machine learning methods while retaining valid statistical inference. Recent research has shown that this is important if correct model specification is difficult, such as when dealing with complex longitudinal data, potentially of small sample size, where relationships and interactions are most likely highly nonlinear and where the number of variables is large compared to the sample size (\citealp{Tran:2019, Schomaker:2019}).

Using causal inference in economics has a long history, starting with path analyses and potential outcome language (\citealp{tinbergen1928supplycurves, wright1934path}) and continuing with regression discontinuity analyses (\citealp{Hahn2008regressiondiscontinuity}), instrumental variable designs (\citealp{imbens2014instrumentalvariables}), and propensity score approaches in the context of the potential outcome framework (\citealp{rosenbaum1983propensity}), among many other methods. More recently, there have been works advocating the use of doubly robust techniques in econometrics (\citealp{Chernozhukov:2018}). From the perspective of statistical inference, this is a very promising suggestion because the integration of modern machine learning methods in causal effect estimation is almost inevitable in areas with a large number of covariates and complex data-generating processes (\citealp{Schomaker:2019}).

However, the application of doubly robust effect estimation can be challenging for (macro-)economic data. First, the causal model that summarizes the knowledge about the data-generating process is often more complex for economic than for epidemiological questions, where most successful implementations have been demonstrated thus far (\citealp{Kreif:2017, Decker:2014, Schnitzer:2014, Schnitzer:2014b, Schnitzer:2016, Tran:2016, Schomaker:2019, Bell-Gorrod:2019}). The task of representing the causal model in a directed acyclic graph (DAG) becomes particularly challenging when considering how economic variables interact with each other over time. Thus, to build a DAG, a thorough review of a vast amount of literature is needed, and economic feedback loops need to be incorporated appropriately. \citet{Imbens:2019}, who discusses different schools of causal inference and their use in statistics and econometrics, as well as different estimation techniques, emphasizes this point:

\hspace*{0.05\textwidth}\parbox{0.95\textwidth}{
{\it{\glqq [...] a major challenge in causal inference is coming up with the causal model.\grqq}}}

Second, even if a causal model has been developed, identification of an estimand has been established and data have been collected, statistical estimation may be nontrivial given the complexity of a particular data set (\citealp{Schomaker:2019}). If the sample size is small, potentially smaller than the number of (time-varying) covariates, recommended estimation techniques can fail, and the development of an appropriate set of learning and screening algorithms is important. The benefits of LTMLE, which is doubly robust effect estimation in conjunction with machine learning to reduce the chance of model misspecification, can be best utilized under a good and broad selection of learners that are tailored to the problem of interest.

Estimating the effect of CBI on inflation is a typical example of a causal inference question that faces all of the challenges described above. Our paper makes five novel contributions to the literature. i) We discuss identification and estimation for our question of interest and estimate the effect of CBI on inflation; ii) we develop a causal model that can be applied to other macroeconomic questions; iii) we demonstrate that it is possible to develop a DAG for economic questions, which is important, as it has been argued that \textit{"the lack of adoption in economics is that the DAG literature has not shown much evidence of the benefits for empirical practice in settings that are important in economics."} \citep{Imbens:2019}; iv) we demonstrate how to integrate machine learning into complex causal effect estimation, including how to define a successful learner set when the number of covariates is larger than the sample size and when there is time-dependent confounding with treatment-confounder feedback \citep{hernan2017causal}; and v) we use simulation studies to study the performance of doubly robust estimation techniques under the challenges described above.

This paper is structured as follows. In the next section, we motivate our question of interest, and this is followed by the general description of our framework. Section \ref{sec:data_analysis} contains the data analysis and describes the doubly robust estimation strategy to estimate the effect of CBI on inflation. In Section \ref{sec:simulations}, we conduct simulation studies motivated by our data analysis. Section \ref{sec:conclusions} concludes the paper.

\iffalse
\begin{figure}[H]

\begin{center}
    \includegraphics[width=\textwidth]{DescriptivePlots/FollowedRegimesOverTime.pdf}
\end{center}
\caption[Share of countries following each intervention]{Share of subjects (countries) that has followed the desired regimes in the past until year t}
    \label{fig:FollowedRegimesOverTime}

\end{figure}

\noindent
See \textit{\href{http://amstat.tfjournals.com/asa-style-guide/}{JASA Style Guide}}. \\
You may cite like this (direct way in case of one or two authors): \cite{ContiuousLongiGformula}.\\
Or like this (direct way in case of more than two authors): \cite{IPTWLongiRegime}.\\
Or like this (direct way with page numbers): \citet[pp. 115-120]{IPTWLongiRegime}.\\
Or like this (indirect way): \citep{IPTWLongiRegime}.\\
Figure \ref{fig:FollowedRegimesOverTime} displays \dots
%%EDITOR'S NOTE: Please ensure that all of the text that is serving as a placeholder is appropriately replaced or removed.

\begin{equation}
    \mathbb{E}(Y_{T+2}^{\overline{d}_{T}}) = \mathbb{E}(\ldots\mathbb{E}(Y_{T+2}|\boldsymbol{\overline{L}_{T+1}},\overline{A}_{T}=\overline{d}_{T},\overline{V}_{1998})\ldots|\boldsymbol{\overline{L}_{1999}},A_{1998}=d_{1998},\overline{V}_{1998})
    \label{eq:seqGFormula}
\end{equation}
\fi

\section{Motivating Question: Central Bank Independence and Inflation}\label{sec:motivation}

When governments have discretionary control over monetary instruments, typically a short-term interest rate, they can prioritize other policy goals over price stability. For instance, after nominal wages have been negotiated (or nominal bonds purchased), politicians may be tempted to create inflation to boost employment and output (gross domestic product, GDP) or to devalue government debt. This is referred to as the time-inconsistency problem of commitments to price stability. It results in an inflation rate higher than what is socially desirable. To overcome this outcome, the literature discusses a variety of commitment mechanisms (also called ``commitment technologies''), ranging from simple rules (such as the imposition of strict rules on the rate of monetary expansion, inflation targeting and nominal exchange rate targeting), contracts between the government and the central bank, reputational forces and, from a practical perspective the best known and implemented mechanism, the delegation of monetary policy-making to an independent central bank. In particular, \citet{rogoff1985optimal} has proposed delegating monetary policy to an independent and ``conservative'' central banker to reduce the tendency to produce high inflation. Here, conservative means that the central banker dislikes inflation more than the government, in the sense that they places a greater weight on price stability than the government does. Once central bankers are insulated from political pressures, commitments to price stability can be credible, which helps to maintain low inflation. Rogoff's seminal paper had a twofold effect: stimulating the implementation of central bank reforms on the policy side and creating avenues for the design of indices that are suitable to capture the degree of independence of these institutions on the research side.

Following these ideas, a considerable policy consensus grew around the potential of having independent central banks to promote inflation stability \citep{bernhard2002political, kern2019imf}. Numerous countries followed this policy advice. Between 1985 and 2012, and excluding the creation of regional central banks, there were 266 reforms to the statutory independence of central banks, 236 of which were being implemented in developing countries. Most of these reforms (77\%) strengthened CBI \citep{garriga2016central}, though some also weakened it. For instance, the law governing the Reserve Bank of Australia was changed in 2002. While previously the governor and board members were appointed by the governor general; in 2002 appointment power was given to the treasurer, which produced a lower independence score. Moreover, whereas board members had been appointed for exactly five years before this amendment, after the amendment the term was specified as not exceeding five years at the discretion of the appointing person \citep{dincer2014central}.

\iffalse
However, while a considerable body of work has investigated the consequences of assigning more independence to monetary policy authorities, the causes of reforms in central banks have received less attention. As pointed out by \citet{masciandaroRomelli.2019}, the large majority of empirical studies essentially consider CBI as an exogenous (independent) variable that can be useful to explain macroeconomic trends. Yet successive research has argued that political institutions such as central banks evolve endogenously as a response to a set of macroeconomic factors \citep{farvaque.2002, aghion2004endogenous, polillo2005globalization, brumm2006effect, brumm2011inflation, bodea2015price, romelli.2018, masciandaroRomelli.2019}. The step forward in this line of research is to consider the degree of CBI as an endogenous (dependent) variable that has to be explained.
\fi

Despite the broad impact of the policy advice to make central banks more independent, the empirical evidence in support of it remains controversial. We investigate the effect of CBI on inflation with a causal framework that treats countries as units in a longitudinal (panel) setup\iffalse, taking the exogenous forces that drive CBI into account\fi. The data set we use in our analysis was created specifically for this purpose and extends the data set from \citet{BRV_Inflation}. To describe and address relevant confounding structures, the crucial question is: What are possible reasons that motivate the decision of a country to adopt a certain degree of CBI? What macroeconomic factors drive central bank independence \citep{farvaque.2002, aghion2004endogenous, polillo2005globalization, brumm2006effect, brumm2011inflation, bodea2015price, romelli.2018, masciandaroRomelli.2019}? Four arguments stand out:

\begin{enumerate}[i)]
\item \textit{Political institutions:} Federally organized countries with good checks and balances grant their monetary institutions greater autonomy and thus a greater level of central bank independence \citep{deHaan.95, Moser.1999, farvaque.2002}.

\item \textit{Political instability:} Central bank reforms are more likely to follow elections, which lead to political consolidation or to changes in the political orientation of the government \citep{romelli.2018}. \citet{cukierman.1995} find that de facto CBI, as measured by the turnover rate of the central bank governor, is lower in less stable political systems.

\item \textit{Past inflation:} \citet{crowe2008central} show that over the period 1990–2003, greater changes in CBI have occurred in countries originally characterized by lower levels of independence and higher inflation. This finding is strengthened by the research of \citet{masciandaroRomelli.2019} where it is shown that countries which experienced long periods of inflation are characterized by a higher inflation aversion, which may cause the government to grant a higher level of CBI. According to \citet{wachtel.2020} the arguments in favor of an independent central bank began to crystallize in the 1980s after a decade or more of traumatic inflationary experience that put a spotlight on central bank policymaking and its failures.

\item \textit{International pressure:} Binding agreements with international money lenders like the International Monetary Fund or the World Bank often require countries to commit to a particular set of policies \citep{blejer2002inflation, gutierrez2003inflation, polillo2005globalization, rodrik2006goodbye, romelli.2018, kern2019imf, reinsberg2020bad}. According to \citet{dincer2014central}, countries with less developed financial markets, more open economies and countries that have participated in IMF programs have more independent central banks. Similarly, \citet{romelli.2018} finds that countries receiving an IMF loan or becoming a member of a currency union adopt reforms that increase CBI. Another type of external pressure can come from regional clustering, which is often found to be cohesive of certain types of reform processes such as democratisations and economic liberalisations \citep{simmons2004globalization, elhorst2013impact, giuliano2013democracy, acemoglu2019democracy}.
\end{enumerate}

Those arguments inform our causal model and estimation strategies in Section \ref{sec:data_analysis}.

\section{Methodological Framework}\label{sec:methods}

\subsection{Notation}\label{sec:methods_notation}

We consider panel data with $n$ units (i.e., countries in our case) studied over time ($t=0,1,\ldots,T$). At each time point $t$, we observe an outcome $Y_t$, an intervention of interest $A_t$ and several time-dependent covariates $L^j_t$, $j=1,\ldots,q$, collected in a set $\mathbf{L}_t=\{L^1_t,\ldots,L^q_t\}$. Variables measured at the first time point ($t=0$) are denoted as $\mathbf{L_0}=\{L^1_0,\ldots,L^{q_0}_0\}$ and are called ``baseline variables''. The intervention and covariate histories of a unit $i$ (up to and including time $t$) are $\bar{A}_{t,i}=(A_{0,i},\ldots,A_{t,i})$ and $\bar{L}^s_{t,i}=(L^s_{0,i},\ldots,L^s_{t,i})$, $s=1,...,q$, respectively, with $q,q_0 \in \mathbb N$.

We are interested in the counterfactual outcome $Y_{t,i}^{\bar{a}_{t}}$ that would have been observed at time $t$ if unit $i \in \{1,\ldots,n\}$ had received, possibly contrary to the fact, the intervention history $\bar{A}_{t,i}=\bar{a}_t$. For a given intervention $\bar{A}_{t,i}=\bar{a}_t$, the counterfactual covariates are denoted as $\bar{\mathbf{L}}_{t,i}^{\bar{a}_{t}}$. If an intervention depends on covariates, it is dynamic. A dynamic intervention ${d}_t(\mathbf{\bar{L}}_{t})=\bar{d}_t$ assigns treatment ${A}_{t,i} \in \{0,1\}$ as a function $\mathbf{\bar{L}}_{t,i}$. If $\mathbf{\bar{L}}_{t,i}$ is the empty set, the treatment $\bar{d}_t$ is static. We use the notation $\bar{A}_t = \bar{d}_t$ to refer to the intervention history up to and including time $t$ for a given rule $\bar{d}_t$. The counterfactual outcome at time $t$ related to a dynamic rule $\bar{d}_t$ is $Y_{t,i}^{\bar{d}_t}$, and the counterfactual covariates at the respective time point are $\bar{\mathbf{L}}_{t,i}^{\bar{d}_t}$. More specific notation concerning the data analysis is given in Section \ref{sec:data_analysis}.

\subsection{Likelihood}\label{sec:methods_likelihood}
If we assume a time ordering of $\mathbf{L}_t \rightarrow A_t$ at each time point, use $Y_T$ as the outcome, and define $Y_t$, $t<T$, to be contained in $\mathbf{L}_t$, the data can be represented as $n$ iid copies of the following longitudinal data structure:
\begin{eqnarray*}
O = (\mathbf{L}_0,A_0,\mathbf{L}_1,A_1,\ldots,\mathbf{L}_{T-1},A_{T-1},Y_{T}) \stackrel{iid}{\sim} P_0
\end{eqnarray*}
Note that in Section \ref{sec:data_DAG}, in the data analysis, the ordering of variables is different. However, for the given ordering, we can write the respective likelihood $\mathcal{L}(O)$ as
\begin{eqnarray*}
p_0(O_i) &=& p_0(\mathbf{L}_{0,i},A_{0,i},\mathbf{L}_{1,i},A_{1,i},\ldots,\mathbf{L}_{T-1,i},A_{T-1,i},Y_{T,i})\\
 &=& p_0(Y_{T,i}|\bar{{A}}_{T-1,i},\bar{\mathbf{L}}_{T-1,i}) \times p_0(A_{T-1}|\bar{\mathbf{L}}_{T-1,i},\bar{{A}}_{T-2,i})   \\
  && \times p_0(\mathbf{L}_{T-1}|\bar{{A}}_{T-2,i},\bar{\mathbf{L}}_{T-2,i})\times \ldots \times p_0(\mathbf{L}_{0,i})\\\
  &=& p_0(Y_{T,i}|\bar{{A}}_{T-1,i},\bar{\mathbf{L}}_{T-1,i})\\
 && \times \left[\prod_{t=0}^{T-1} \underbrace{p_0({A}_{t,i}|\bar{\mathbf{L}}_{t,i},\bar{A}_{t-1,i})}_{g_{0,A_t}} \right] \times \left[\prod_{t=0}^{T-1} \underbrace{p_0(\mathbf{L}_{t,i}|\bar{A}_{t-1,i},\bar{\mathbf{L}}_{t-1,i})}_{\tilde{q}_{0,\mathbf{L}_t}} \right]  \,.\\
\end{eqnarray*}
In the above factorization, $p_0(\cdot)$ refers to the density of $P_0$ (with respect to some dominating measure) and $A_{-1} := \mathbf{L}_{-1} := \emptyset$. If an order for $\mathbf{L}_t$ is given, e.g., $L^1_t \rightarrow \ldots \rightarrow L^q_t$, a more refined factorization is possible. In line with the notation of other papers (e.g., \citealp{Tran:2019}), we define the \textit{$q$-portion} of the likelihood to also contain the outcome: ${q}_{0,\mathbf{L}_t} := \tilde{q}_{0,\mathbf{L}_t} \times p_0(Y_{T,i}|\bar{A}_{T-1,i},\bar{\mathbf{L}}_{T-1,i})$. Similarly, we define $g_0 := \prod_{t=0}^T g_{0,A_t}$ and $q_0 := \prod_{t=0}^{T} {q}_{0,\mathbf{L}_t}$.

\subsection{On the distinction between the causal and statistical model}
Estimating causal effects cannot be established from data alone but requires additional structural (i.e. causal) assumptions about the data\--gen\-erating process. Therefore, any causal analysis comes with both a structural (i.e. causal) and a statistical model. The former can be represented by a directed acyclic graph (DAG), which encodes conditional independence assumptions and is logically equivalent to a (non-parametric) structural equation framework. Ideally, the structural model is supported by knowledge from the literature. The statistical model encodes assumptions about the family of possible observed data distributions associated with the DAG, with the ultimate aim to estimate post-intervention distributions and quantities. With doubly robust effect estimation, any parametric assumptions are typically eschewed to avoid model mis-specification; and to incorporate machine learning while retaining valid inference. In our framework and analyses below, we proceed as follows: for the causal model, we begin with the basic assumption that variables can be affected by the past, but not the future (Section \ref{sec:methods_causal_model}). In our analysis in Section \ref{sec:data_data}, we then make more detailed assumptions with respect to the causal model: we encode our structural assumptions in a DAG (Figure \ref{fig:DAG}) and support this model with references from the economic literature (Appendix). For the statistical model, we first don't impose any parametric restrictions on the statistical model (Section \ref{sec:methods_stat_model}). In the analysis (Section \ref{sec:data_data}), we then use the above likelihood factorization and targeted maximum likelihood estimation with super learning, to avoid any overly restrictive parametric assumptions.

\subsection{Statistical Model}\label{sec:methods_stat_model}
In line with the notation of Section \ref{sec:methods_likelihood}, we consider a statistical model $\mathcal{M}=\{P = q \times g: q \in \mathcal{Q}, g \in \mathcal{G}\}$ for the true distribution $P_0$ that requires minimal (parametric) assumptions. In contrast to many medical applications, we do not impose restrictions on this model; that is, $A_t$ and $Y_t$ are not deterministically determined for any given data history. Once an intervention is implemented, it can be stopped at any time point and potentially started again. Similarly, the outcome can be observed at any time point, and we do not assume that censoring is possible.

\subsection{Causal Model}\label{sec:methods_causal_model}
Causal assumptions about the data-generating process are encoded in the model $\mathcal{M}^{\mathcal{F}}$. This nonparametric (structural equation) model states our assumptions about the time ordering of the data and the causal mechanism that gave rise to the data. Thus far, it relates to
\begin{eqnarray*}
Y_T &=& f_{Y_T}(\bar{A}_{T-1},\bar{\mathbf{L}}_{T-1},U_{Y_T})\\
\boldsymbol{L_t} &=& f_{\boldsymbol{L_t}}(\bar{A}_{t-1},\bar{\boldsymbol{L}}_{t-1},\boldsymbol{U_{L_{t}}}): t=0,1,\ldots,T-1\\
A_t &=& f_{A_t}(\bar{\boldsymbol{L}}_{t},\bar{A}_{t-1},U_{A_t}): t=0,1,\ldots,T-1
\end{eqnarray*}
where $\mathbf{U}:=(U_{Y_T},\boldsymbol{U_{\boldsymbol{L_{t}}}},U_{A_t})$ are unmeasured variables from some underlying distribution $P_{\boldsymbol{U}}$. For now, we do not make any assumptions regarding $P_{\boldsymbol{U}}$. However, in the data example further below, we need to enforce some restrictions on this distribution. The functions $f_{O}(\cdot)$ are (deterministic) nonparametric structural equations that assume that each variable may be affected only by variables measured in the past and not those that are measured in the future. Section \ref{sec:data_DAG} refines the causal model for the data-generating process of the motivating question and represents any additional assumptions made in a DAG.

\subsection{Causal Target Parameter and Identifiability}\label{sec:methods_target_parameter}
In this paper, we focus on the differences in intervention-specific means, i.e., in target parameters such as
\begin{eqnarray}\label{eqn:target_parameter}
\psi_{j,k} = \mathbb{E}(Y_{T}^{\bar{d}_{t}^{j}})-\mathbb{E}(Y_{T}^{\bar{d}_t^{k}}) , \quad j\neq k\,.
\end{eqnarray}
If we set the intervention according to a static or dynamic rule ($\bar{A}_t=\bar{d}_t^l$ $\forall t$) with $l \in \{j,k\}$ in the causal model $\mathcal{M}^{\mathcal{F}}$, we obtain the post-intervention distribution $P_0^{\bar{d}_t^l}$. The counterfactual outcome $Y_{T}^{\bar{d}_t^l}$ is the one that would have been observed had $A_t$ been set deterministically to $0$ or $1$ according to rule $\bar{d}_t^l$. We thus restrict the set of possible interventions to those where the intervention is binary $A_{t,i} \in \{0,1\}$.

It has been shown that target parameters of the form (\ref{eqn:target_parameter}) can be identified under the (partly untestable) assumptions of consistency, conditional exchangeability and positivity, which are defined below. Specifically, it follows from the work of \citet{Bang:2005} and \cite{vanderLaan:2012} that given these three assumptions, using the iterative conditional expectation rule, and for the particular time-ordering as defined in Section \ref{sec:methods_likelihood}, we can write the target parameter as

\begin{eqnarray}\label{eqn:seq_g_formula}
\psi_{j,k} &=& \mathbb{E}(Y_{T}^{\bar{d}_t^j})-\mathbb{E}(Y_{T}^{\bar{d}_t^k}) \nonumber\\
&=& \mathbb{E}(\,\mathbb{E}(\,\ldots\mathbb{E}(\,\mathbb{E}(Y_T|\bar{\mathbf{A}}_{T-1}=\bar{d}^j_{T-1}, \mathbf{\bar{L}}_{T-1}) | \bar{\mathbf{A}}_{T-2}=\bar{d}^j_{T-2}, \mathbf{\bar{L}}_{T-2}\,)\ldots|\bar{A}_{0}=\bar{d}^j_{0}, \mathbf{{L}}_{0}\,)|\mathbf{{L}}_{0}\,)\,) - \nonumber\\
 && \mathbb{E}(\,\mathbb{E}(\,\ldots\mathbb{E}(\,\mathbb{E}(Y_T|\bar{\mathbf{A}}_{T-1}=\bar{d}^k_{T-1}, \mathbf{\bar{L}}_{T-1}) | \bar{\mathbf{A}}_{T-2}=\bar{d}^k_{T-2}, \mathbf{\bar{L}}_{T-2}\,)\ldots|\bar{A}_{0}=\bar{d}^k_{0}, \mathbf{{L}}_{0}\,)|\mathbf{{L}}_{0}\,)\,) \,.
\end{eqnarray}

The assumptions of consistency, conditional exchangeability and positivity have been discussed in the literature in detail (\citealp{Daniel:2011, Daniel:2013, Robins:2009, Young:2011, Tran:2019}). Briefly, consistency is the requirement that $Y^{\bar{d}_t}_T = Y_T$ if $\bar{\mathbf{A}}_{t-1} = \bar{d}_{t-1}$ and $\bar{\mathbf{L}}_{t}^{\bar{d}_{t}}=\bar{\mathbf{L}}_{t}$ if $\bar{\mathbf{A}}_{t-1} = \bar{d}_{t-1}$. Conditional exchangeability requires the counterfactual outcome under the assigned treatment rule to be independent of the observed treatment assignment, given the observed past: $Y^{\bar{d}_t}_T\coprod {\mathbf{A}_{t-1}|\bar{\mathbf{L}}_{t-1}, \bar{\mathbf{A}}_{t-2}}$ $\forall \bar{\mathbf{A}}_t=\bar{d}_t, \bar{\mathbf{L}}_t=\bar{\mathbf{l}}_t, \forall t$, and positivity says that each unit should have a positive probability of continuing to receive the intervention according to the assigned treatment rule, given that this has been done so far, and irrespective of the covariate history: $P(\mathbf{A}_t=\bar{d}_t|\bar{\mathbf{L}}_t=\bar{\mathbf{l}}_t,\bar{\mathbf{A}}_{t-1}=\bar{d}_{t-1})>0$  $\forall t,\bar{d}_t,\bar{\mathbf{l}}_t$ with $P(\bar{\mathbf{L}}_t=\bar{l}_t,\bar{\mathbf{A}}_{t-1}=\bar{d}_{t-1}) \neq 0$.

In principle, (conditional) exchangeability can be evaluated graphically for an assumed structural model represented in a DAG using the back-door criterion (\citealp{Pearl:2010, Molina:2014}); i.e., by closing all back-door paths and by nonconditioning on descendants of the intervention. For multiple time-point interventions, a generalized version of this criterion can be used to verify conditional exchangeability. This requires blocking all back-door paths from $A_t$ to $Y_T$ that do not go through any future treatment node $A_{t+1}$ (\citealp{hernan2017causal}). More generally, it has been suggested to use single-world intervention graphs to verify exchangeability, particularly to evaluate identification for complex dynamic interventions. See Richardson and Robins for details (\citealp{Richardson:2013}).

\subsection{Effect estimation with Longitudinal TMLE}\label{sec:methods_estimation_LTMLE}
The longitudinal TMLE estimator (\citealp{vanderLaan:2012}) relies on equation (\ref{eqn:seq_g_formula}). To estimate $\psi_{j,k}$, one can separately evaluate each of the two nested expectation terms and integrate out $\bar{\mathbf{L}}_{T-1}$ with respect to the post-intervention distribution $P_0^{\bar{d}_t^l}$. To improve inference with respect to $\psi_{j,k}$, a targeted estimation step at each time point yields a doubly robust estimator of the desired target quantity (see \citet{vanderLaan:2011} or \citet{Schnitzer:2017} for details). Specifically, we recur to the following algorithm for $t=T,...,1$:
\begin{enumerate}
\item Estimate $\bar{Q}_T = \mathbb{E}(Y_T|\mathbf{\bar{A}}_{T-1}, \mathbf{\bar{L}}_{T-1})$ with an appropriate model (for $t=T$). If $t<T$, use the prediction from step 3d (of iteration $t-1$) as the outcome, and fit the respective model. The estimated model is denoted as $\hat{{Q}}_{0,t}$.
\item Now, plug in $\mathbf{\bar{A}}_{t-1}={\bar{d}}^l_{t-1}$ based on rule $\bar{d}_t^l$, and use the fitted model from step 1 to predict the outcome at time $t$ (which we denote as $\hat{Q}^{\bar{d}_t^l}_{0,t}$).
\item To improve estimation with respect to the target parameter, update the initial estimate of step 2 by means of the following regression:
\begin{enumerate}[a)]
\item The outcome refers again to the measured outcome for $t=T$ and to the prediction from item 3d (of iteration $t-1$) if $t<T$.
\item The offset is the original predicted outcome $\hat{Q}^{\bar{d}_t^l}_{0,t}$ from step 2 (iteration $t$).
\item The ``clever covariate'' is defined as:
\begin{eqnarray}\label{eqn:clever_covariate}
     {H}_{t-1} = \prod_{s=0}^{t-1} \frac{I(\bar{{A}}_s=\bar{d}_{s})}{g_{0,A_t=\bar{d}_s^l}}
     \end{eqnarray}
with $g_{0,A_t=\bar{d}_s^l}=P(A_s=\bar{d}_s^l|\bar{\mathbf{L}}=\bar{\mathbf{l}}_s,\bar{A}_{s-1}=\bar{d}_{s-1}^l)$. The estimate of $g_{0,A_t=\bar{d}_s^l}$ is denoted as $\hat{g}_{A_t=\bar{d}_s^l}$.
\item predict the updated (nested) outcome, $\hat{Q}_{1,t}^{\bar{d}_t^l}$, based on the model defined through 3a, 3b, and 3c.
\end{enumerate}
This model contains no intercept. Alternatively, the same model can be fitted with $H_{t-1}$ as a weight rather than a covariate \citep{Kreif:2017, Tran:2019}. In this case, an intercept is required. We follow the latter approach in our implementations.
%\end{enumerate}
%\begin{enumerate}
%\setcounter{enumi}{3}
\item The estimate for $\mathbb{E}(Y_{T}^{\bar{d}_t^l})$ is obtained by calculating the mean of the predicted outcome from step 3d (where $t=1$).
\item Confidence intervals can, for example, be obtained using the vector of the estimated influence curve; see \citet{Tran:2018} for a review of adequate choices.
\item Repeat 1.-5. to estimate $\mathbb{E}(Y_{T}^{\bar{d}_t^j})$ and $\mathbb{E}(Y_{T}^{\bar{d}_t^k})$. Now, $\hat{\psi}_{j,k}$ and its corresponding confidence intervals can be calculated.
\end{enumerate}

\subsubsection{Inference and Properties of LTMLE}\label{sec:methods_properties}

For an arbitrary distribution $P\in \mathcal{M}$ and a specific intervention rule $g = g(P)$ we consider the statistical model $M(g) = \{P^* \in \mathcal{M} : g(P^*) = g\}$ for the respective treatment rules $g$. With such a model we could estimate $\psi
^*$ with the algorithm described in \ref{sec:methods_estimation_LTMLE}. For $\psi^*$ it can be shown (e.g. \citealp{van2018targeted}) that
%\begin{equation}
%    \sqrt{n}(\hat{\bm{\psi^*}}- \bm{\psi^*}) \xrightarrow{d} N(\bm{0},\bm{\Sigma}^*)
%\end{equation}
$\hat{\psi}^*$ is an asymptotically efficient estimator of $\psi^*$ where
\begin{equation}
    \sqrt{n}(\hat{\psi}
^* - \psi^*) \xrightarrow{d} N(0,\sigma^{2,\ast}).
\end{equation}
The variance can be estimated with the sample variance of the estimated influence curve. This is essentially because the construction of the covariate in step 3c, guarantees that the estimating equation corresponding to the (efficient) influence curve is solved, which in turn yields desirable (asymptotic) inferential properties. The influence curve emerges from the linear span of the scores (i.e. first derivative) of the logistic loss for the density of the outcome variable (evaluated at zero) for a given value of the clever covariate \citep{Schnitzer:2014b}. Thus, in the longitudinal case, for interventions rules $\bar{g}_{t}$, these score components can be summed across the points in time which yields the efficient influence curve
%$\hat{\bm{\psi^*}}$ is asymptotically linear with a vector-valued influence curve $\bm{IC}^*$. We can estimate $\bm{\Sigma}^*$ consistently with the empirical covariance matrix
%\begin{equation}
%    \hat{\bm{\Sigma}}^* = P_n(\hat{\bm{IC}^*}) P_n(\hat{\bm{IC}^*})^\top
%\end{equation}
%where $P_n$ denotes the empirical distribution. The influence curve can be estimated through
%$\hat{\bm{IC}^*} = (\hat{IC}^*_{g_{1}},\ldots, \hat{IC}^*_{g_{k}})^\top$
\begin{equation}
\hat{IC}^* = \left\{\sum_{t=1}^{T} \hat{H}_{t-1}^{*} [{\hat{Y}_t}^{* \ \bar{d}_{t} = \bar{g}_{t}} - {\hat{Y}_{t-1}}^{* \ \bar{d}_{t-1} = \bar{g}_{t-1}}] \right\} + \hat{Y}_{0}^{* \ \bar{d}_{t} = \bar{g}_{t}} - \hat{\psi}^*.
\end{equation}

\subsubsection{Data-Adaptive Estimation for Complex (Macroeconomic) Data}\label{sec:methods_data_adaptive}
The above estimation procedure is doubly robust, which means that the estimator is consistent as long as either the Q- or the g-models (steps 1 and 3c in the algorithm described above) are estimated consistently (\citealp{Bang:2005}). If both are estimated consistently (at reasonable rates), the estimator is asymptotically efficient because the construction of the covariate in step 3c guarantees that the estimating equation corresponding to the efficient influence curve is solved, which in turn yields desirable (asymptotic) inferential properties (\citealp{vanderLaan:2011, Schnitzer:2017}).

To estimate the conditional expectations in the algorithm, one could use (parametric) regression models. Under the assumption that they are correctly specified, this approach would be valid. However, in the context of complex macroeconomic data, as in our motivating example below, it is challenging to estimate appropriate parametric models because of small sample sizes, a large number of relevant variables and complex nonlinear relationships. Longitudinal TMLE can (in contrast to many competing estimation techniques) incorporate machine learning algorithms while still retaining valid inference to reduce the possibility of model misspecification. However, in the settings presented below, machine learning approaches need to be tailored to the specific problem and address the following challenges:

\begin{enumerate}[i)]
\item \textbf{Complexity}: Macroeconomic relationships are often highly nonlinear and have various interactions of higher order, which need to be modeled in a sophisticated manner while taking into account the time ordering of the data.
\item \textbf{Dispensable variables}: The inclusion of covariates in the estimation procedure that are not required for identification, i.e., do not block any back-door paths, can potentially be harmful even if they are not colliders or mediators \citep{schnitzer2016variable}; that is, the inclusion of such variables can increase the finite-sample variance and lead to small estimated probabilities of following a particular treatment rule given the past, which may be both incorrectly interpreted as positivity violations and make the updating step in the TMLE algorithm unstable. They may also amplify bias \citep{Pearl:2011}.
\item \textbf{p$\bm{>}$n}: For longitudinal macroeconomic data, the number of parameters is often larger than the sample size. This is because for long follow-up, the whole covariate history needs to be considered, interactions may be nonlinear, and different variables may have different scales and features that need to be modeled adequately. Consequently, one needs to either reduce the number of parameters with an appropriate estimation procedure or eliminate variables beforehand using variable screening. It has been argued that screening of variables is inevitable to facilitate estimation with LTMLE in many settings \citep{schnitzer2016variable}.
\end{enumerate}

Section \ref{sec:data_estimation_LTMLE} recommends possible approaches to tackle these challenges in common macroeconomic settings.

\section{Data Analysis: Estimating the Effect of Central Bank Independence on Inflation}\label{sec:data_analysis}

\subsection{Data}\label{sec:data_data}
%%EDITOR'S NOTE: Abbreviations and acronyms are typically defined the first time the term is used within the main text and then used throughout the remainder of the manuscript. Please consider adhering to this convention. The target journal may have a list of abbreviations that are considered common enough that they do not need to be defined.

We accessed databases of the World Bank and the International Monetary Fund to collect annual data for economic, political and institutional variables. Our outcome of interest is inflation in 2010 ($Y_{2010}$). All covariates are measured annually at equidistant points in time for $t^{\ast}=1998,\ldots,2010$. The intervention variable is central bank independence at time $t^{\ast}$ (CBI, $A_{t^{\ast}}$), which we define as suggested by \citet{dincer2014central}: their CBI index measures several dimensions of independence and runs from 0, the lowest level of independence, to 1, the highest level of independence. It contains considerations such as the independence of the chief executive officer (CEO) and limits on his/her reappointment, the bank's independence in terms of policy formulation, its objective or mandate, the stringency of limits on lending money to the public sector, measures of provisions affecting (re)appointment of board members other than the CEO, restrictions on government representation on the board, and intervention of the government in exchange rate policy formulation. This definition implies that our central bank independence index is an intervention that can in principle be modified through legislative amendments, although it is the very nature of an index to represent multiple facets of a phenomenon that cannot not be easily dealt with in an actual experiment. We binarized \citeauthor{dincer2014central}s' index at a value of 0.45 by setting countries with a value greater than 0.45 to 1 (independent) for each time point and 0 (dependent) otherwise.   We  then  used  the  binarized  index  for  estimation.   The  trajectories  of  their  original  indices and our binarized version can be seen in Figure \ref{fig:CBI_regime_switches}. Our outcome variable is defined as the year-on-year changes (expressed as annual percentages) of average consumer prices measured by a consumer price index (CPI). A CPI measures changes in the prices of goods and services that households consume. To calculate CPIs, government agencies conduct household surveys to identify a basket of commonly purchased items and then track the cost of purchasing this basket over time. The cost of this basket at a given time, expressed relative to a base year, is the CPI, and the percentage change in the CPI over a certain period is referred to as consumer price inflation, the most widely used measure of inflation. Our measured covariates are $\mathbf{L}_{t^{\ast}}=\{L^1_{t^{\ast}},\ldots,L^{18}_{t^{\ast}}\}$ and include a variety of macroeconomic variables such as money supply, energy prices, economic openness, institutional variables such as central bank transparency and monetary policy strategies, and political variables (see Figure \ref{fig:DAG}, Table \ref{tab:VerticesDescription} and \citet{BRV_Inflation} for details.). In line with the notation of Section \ref{sec:methods}, we consider $Y_{t^{\ast}}$, $t^{\ast}<T=2010$, to be part of $\mathbf{L}_{t^{\ast}}$, i.e., we define $L^8_{t^{\ast}} := Y_{t^{\ast}}$.

Our aim was to include as many countries as possible in our analysis. This entailed a tradeoff between the number of countries and the completeness of the data set. We were able to collect annual data from 1998 to 2010 for 124 countries for 14 explanatory variables and for the dependent variable $Y_{t^{\ast}}$. We further derived growth rates and other indicators from those measured variables to capture data for all 18 covariates ($\mathbf{L}_{t^{\ast}}$). Some of the data were missing, however. To decide whether the missing data were likely missing not at random (MNAR) and therefore possibly not useful without making additional assumptions, we examined countries' characteristics. We decided that observations for certain variables, countries or groups of countries had to be excluded because they were not available; for instance, sometimes wars, insufficiently developed institutions, social unrest or other reasons made the collection of data impossible. We split the data set according to our assessment of whether the observation was MNAR. Data that we regarded as missing at random (MAR) (2.7\% of the data set) were multiply imputed using \texttt{Amelia II} (\citealp{AmeliaII}), taking the time-series cross-sectional structure of the data into account. We did not impute data that were likely MNAR. However, some variables that were categorized as MNAR were used in the analysis (e.g., CBI). As a result, we obtained observations for 60 countries and 13 points in time (i.e., calendar years 1998-2010) for 19 measured variables ($L_{t^{\ast}}^1$,\ldots,$L_{t^{\ast}}^{7}$,$L_{t^{\ast}}^9$,\ldots,$L_{t^{\ast}}^{18}$,$Y_{t^{\ast}} \equiv L^8_{t^{\ast}}$,$A_{t^{\ast}}$). In this final data set, 0.1\% of observations were missing and thus imputed.

According to the World Bank's income classification, approximately 20\% of the remaining 60 countries are low-income countries, 36\% belong to the lower-middle-income category, 27\% to the upper-middle-income category and 17\% belong to the high-income category. While our sample reflects considerable heterogeneity with respect to countries' development level, it is possible that the included countries are not representative of all countries in the world: as many excluded countries faced periods of violent conflicts or had no well-developed governmental institutions, our sample likely reflects economies of (reasonably) stable countries.

\subsection{Target Parameters and Interventions}\label{sec:data_target_parameter}
Our target parameters are ATEs as defined in (\ref{eqn:target_parameter}). To be more specific, consider the following three interventions, of which two are static and one dynamic, each of them applied to $\forall t^\ast \in \{1998,\ldots,2008\}$:

\begin{flalign*}
\bar{d}_{t^\ast}^{1} \phantom{\quad\,\,\bar{L}^1_{t^{\ast}-1}} &= \,\, \left\{a_{t^\ast}=1\right.&
\end{flalign*}
\begin{flalign*}
\bar{d}_{t^\ast,i}^{2} (\bar{L}^8_{t^\ast-1}) &=\left\{ \begin{array}{cl}
               a_{t^\ast,i}=1 & \quad \mbox{if} \quad \widehat{\text{median}}({L^8_{t^\ast-1,i},\ldots,L^8_{t^\ast-7,i}}) \leq 0 \quad \mbox{or} \quad  \widehat{\text{median}}({L^8_{t^\ast-1,i},\ldots,L^8_{t^\ast-7,i}}) \geq 5\\
               a_{t^\ast,i}=0 &  \quad \mbox{otherwise}
               \end{array}
               \right. &
\end{flalign*}
\begin{flalign*}
\bar{d}_{t^\ast}^{3} \phantom{\,\,\quad\bar{L}^1_{t^\ast-1}} &= \,\, \left\{a_{t^\ast}=0\right.&
\end{flalign*}

A country's central bank is set to be either independent (i.e., $\bar{d}_{t^\ast}^{1}$) or dependent (i.e., $\bar{d}_{t^\ast}^{3}$) during the whole time period under the first and third intervention above. This means that we intervene on the first 11 (i.e. from 1998-2008) out of 13 points (i.e. from 1998-2010) in time. This is because we assume a two-year lag between the CBI intervention and its effect on inflation. The transmission mechanism of monetary policy is said to exhibit “long and variable” lags \citep{friedman.1972,batini_nelson.2001,Goodhart.2001}. In line with this view, inflation-targeting central banks have adopted a value between 12 and 24 months as transmission lag (the horizon at which the response of prices becomes the strongest). Theoretical models usually imply transmission lags of similar length \citep{taylor.2012}. According to a meta analysis of 67 published studies for 30 different countries \citep{rusnak.2013} the average transmission lag is 29 months. However, transmission lags are longer in developed economies (25-50 months) than in post-transition economies (10-20 months). Overall, after filtering out effects of misspecifications, the results suggest that prices bottom out approximately two-and-a-half years after a monetary contraction. Given the heterogeneity of our countries, we chose a somewhat shorter lag to take into account countries' differences in their stage of development. 

The second (dynamic) intervention sets a country's central bank to be independent if its median inflation rate in the past 7 years was below 0\% or greater than 5\%. The rationale for this relates to the fact that excessive inflation and deflation over several years are considered to produce harmful effects on a country's economy (see, e.g., \citet{10.2307/1910352, fisher1933debt}). To guarantee price stability, which excludes inflation beyond a certain level and deflation, an independent central bank is required. Over the last twenty years, the optimal level of inflation has been associated with approximately 2\% \citep{mishkin2017rethinking}. If a country's inflation is constantly well above this level, in our case 5\%, it will change the status of its central bank towards independence. The same holds for an inflation rate systematically falling below a value of zero. Note that for the dynamic intervention $\bar{d}_{t^\ast,i}^{2}$, data prior to 1998 had to be collected and utilized.

We define the following two target parameters:
\begin{eqnarray}
\psi_{1,3} = \mathbb{E}(Y_{2010}^{\bar{d}_{t^\ast}^1})-\mathbb{E}(Y_{2010}^{\bar{d}_{t^\ast}^3}) \,, \\
\psi_{2,3} = \mathbb{E}(Y_{2010}^{\bar{d}_{t^\ast}^2})-\mathbb{E}(Y_{2010}^{\bar{d}_{t^\ast}^3}) \,.
\end{eqnarray}
The first, $\psi_{1,3}$, quantifies the expected difference in inflation two years after the last intervention (i.e., in 2010) if every country had an independent central bank for 11 years in a row compared to a dependent central bank for 11 consecutive years. The second, $\psi_{2,3}$, quantifies the effect that would have been observed if every country's central bank had become independent for time points when the country's median inflation in the preceding 7 years had been outside the range from 0 to 5, compared to a strictly dependent central bank for 11 consecutive years (i.e., for the period 1998-2008).

\subsection{Statistical and Causal Model (DAG)}\label{sec:data_DAG}
We separate the measured variables into blocks. The first block comprises $\mathbf{L}^{A}_{t^{\ast}}:=\{L^{1}_{t^{\ast}},\ldots,L^{7}_{t^{\ast}}, L^{9}_{t^{\ast}},\ldots, L^{15}_{t^{\ast}}\}$, and the second comprises $\mathbf{L}^{B}_{t^{\ast}}:=\{L^{16}_{t^{\ast}},\ldots,L^{18}_{t^{\ast}}\}$. In line with Sections \ref{sec:methods_likelihood} and \ref{sec:methods_stat_model}, we do not make any overly restrictive assumptions with respect to our statistical model. First, we assume that our data come from a general true distribution $P_0$ and are ordered such that

\begin{eqnarray*}
     O = (Y_{1998},\mathbf{L}_{1998}^{A},A_{1998},\mathbf{L}_{1998}^{B},Y_{1999},\mathbf{L}_{1999}^{A},A_{1999},\mathbf{L}_{1999}^{B}\ldots,Y_{2009},\mathbf{L}_{2009}^{A},A_{2009},\mathbf{L}_{2009}^{B},Y_{2010}) \stackrel{iid}{\sim} P_0 \,.
\end{eqnarray*}

In the context of our application, we do not need to make any deterministic assumptions regarding our intervention assignment: a central bank can, in principle, be independent or dependent at any point in time, irrespective of the country's history -- and thus be intervened upon.

As discussed in Section \ref{sec:methods_causal_model}, we assume that each variable may be affected only by variables measured in the past and not those that are measured in the future. In addition, we make several assumptions regarding the data-generating process, which are summarized in the DAG in Figure \ref{fig:DAG}. Not all variables listed in $O$ are needed during estimation; see Section \ref{sec:data_estimation_LTMLE}.

The DAG contains both measured variables (in grey colour) and unmeasured variables (in white colour). The outcome variable is coloured in green, and the intervention in red.

The DAG summarizes our knowledge of the transmission channels of monetary policy. An arrow $A \rightarrow B$ reflects our belief, corroborated by economic theory, that $A$ may cause $B$, whereas an absence of such an arrow states that we assume no causal relationship between the respective two variables. Figure \ref{fig:DAG} has been developed based on economic theory. For example, arrow number 6 describes the causal effect from real GDP (Output) on one component of companies' price setting (Price Markup), which is motivated by the fact that changes in demand (c.p.) in the goods market enable companies to set higher prices in a profit-maximizing environment. Detailed definitions of the considered variables, as well as detailed justification for the assumptions encoded in our DAG, are given in Tables \ref{tab:VerticesDescription} and {\ref{tab:Edge Description}} in the Appendix as well as in Section \ref{sec:motivation}.

\subsection{Identifiability Considerations}\label{sec:data_identifiability}

The DAG shows the causal pathways through which CBI can affect consumer prices and thus ultimately inflation. We next explain the main paths from the intervention node to consumer prices. An independent central bank sets its policy tools autonomously to achieve its objective(s). Moreover, an independent central bank is less pressured to pursue an overly expansionary monetary policy that would produce only high inflation. Such a central bank is more likely to live up to its word, which increases its credibility (arrow 74). Higher credibility keeps inflation expectations in check (arrow 32). The more contained inflation expectations are, the lower the demands for nominal wage compensation will be (arrow 75), which, in turn, keeps labor costs (arrow 29), production costs (arrow 23) and companies' prices (arrow 3) low. This will ultimately also be reflected in relatively low consumer prices (arrow 2). Another pathway from the intervention to the outcome acts through monetary policy decisions. Following an intervention, monetary policy makers' time preferences are reduced (arrow 69), and this will be taken into account in their monetary policy decisions (arrow 49). Monetary policy decisions are mirrored in money supply (arrow 52), which is tantamount to banks' loan creation (arrow 66) and, as a result, affects firms' investment decisions (arrow 67) and thus output (arrow 11). The final stage affects firms' markups (arrow 6) in their prices with a final effect on consumer prices (arrows 4 and 2).

\include{DAG}
\restoregeometry

There are several back-door paths from the intervention to the outcome. They all start with arrow 98 because CBI status is ultimately influenced by government decisions, which are in turn affected by past inflation, political institutions and political stability see Section \ref{sec:motivation} or a detailed justification. As an example, consider the back-door path that goes through government decisions (arrow 98) and past inflation (arrow 101): the latter  affects current monetary policy decisions (arrow 65). Monetary policy will in turn impact the formation of inflation expectations (arrow 59) or the money supply (arrow 52). Along edges 66, 67, 11, 6, 4 and 2, this affects the outcome.

Under the assumption that the DAG as motivated in Appendix \ref{sec:appendix_arrows_DAG} is correct, establishing identification in terms of the (generalized) back-door criterion requires the following considerations: all back-door paths start with arrow 98 and can be blocked by conditioning on the following 4 variables: past inflation ($L_{t^{\ast}}^{9}$), central bank transparency ($L_{t^{\ast}}^{13}$), political institution ($L_{t^{\ast}}^{14}$) and political instability ($L_{t^{\ast}}^{15}$). There are various paths from the intervention to the outcome that start with edges 69, 49 and 52. All those paths contain mediators one should \textit{not} necessarily condition on in our example because otherwise the effect of CBI on inflation through these paths would be blocked (\citealp{hernan2017causal}). The same considerations apply to the paths starting with edges 74 and 32.

In summary, our DAG suggests that all back-door paths from $A_{t^\ast}$ to the outcome (that do not go through any future treatment node $A_{t^{\ast}+1}$) can be blocked by including $L_{t^{\ast}}^{9}$, $L_{t^{\ast}}^{13}$, $L_{t^{\ast}}^{14}$ and  $L_{t^{\ast}}^{15}$ in the analysis. As many other variables lie on a mediating path from the intervention to the outcome (i.e., are descendants of $A_{t^{\ast}}$), they should not be conditioned upon.

We argue that the developed DAG should serve as the basis for identification considerations and estimation strategies. However, in complex macroeconomic situations, violations of this causal model need to be taken into account, and other estimation strategies may also be useful. We now explain how this can be facilitated.

\subsection{Data-adaptive Estimation with longitudinal TMLE}\label{sec:data_estimation_LTMLE}
%%EDITOR'S NOTE: Please use a consistent capitalization and punctuation format for section headings throughout the manuscript. Some journals request a specific style, so please review the journal's guidelines.

We can, in principle, follow the algorithm described in Section \ref{sec:methods_estimation_LTMLE} to estimate the target quantity of interest. This includes estimation of the (nested) outcome model $\bar{Q}_{t^\ast}$ (step 1) and the intervention model $g_{0,A_{t^\ast}=\bar{d}^l_s}$ (step 3c) for each time point. That is, we can estimate the $g$-model for $t^\ast=1998, \ldots,2008$ and $Q_{t^{\ast}}$ for $t^\ast = 2000,\ldots,2010$. As mentioned above, the DAG assumes a 2-year lag before an independent central bank can potentially affect the outcome. It is thus sufficient to estimate the first Q-model in 2000 given the assumed lag structure in the DAG. We define $Y_T := Y_{2010}$, which corresponds to the value of inflation in 2010, while $\bar{d}_{t^{\ast}}^{1}$, $\bar{d}_{t^{\ast},i}^{2} (\bar{L}^8_{t^{\ast}-1})$ and $\bar{d}_{t^{\ast}}^{3}$ are the interventions targeting CBI as described in Section \ref{sec:data_target_parameter}.

We consider three approaches to covariate inclusion. The first is based on the identifiability considerations related to our DAG, and the other two refine variable inclusion criteria based on the scenario in which some structural causal assumptions in the DAG may be incorrect.

\begin{enumerate}[i)]
\item \textbf{DAG-based approach} (\textit{PlainDAG}): Based on the identifiability arguments from Section \ref{sec:data_identifiability}, $\mathbf{\bar{L}}_{t^{\ast}}$ contains only the relevant baseline variables from 1998 that were measured prior to the first intervention node, as well as $L_{t^{\ast}}^{9}$, $L_{t^{\ast}}^{13}$, $L_{t^{\ast}}^{14}$ and  $L_{t^{\ast}}^{15}$.
\item \textbf{Greedy super learning approach} (\textit{ScreenLearn}): This approach contains the full set of measured variables $\mathbf{L}_{t^\ast}$. This approach assumes that each variable could potentially lie on a back-door path but that this was undiscovered due to misspecification of the causal model. For example, a researcher who argues that bank loans directly affect a central bank's independence (i.e., that there is an arrow from bank loans to CBI) would have to consider a back-door path along arrows 67, 11, 6, 4, 2 and thus include public debt in $\mathbf{L}_{t^\ast}$. Similarly, if it is doubted that some variables are not necessarily mediators but rather confounders on a back-door path that exists due to unmeasured variables, e.g., $\text{\textit{CBI}} \leftarrow \text{\textit{unmeasured variable}} \rightarrow \text{\textit{Output}} \rightarrow \ldots \rightarrow \text{\textit{Consumer Prices}}$, then measured variables such as \textit{Output} (real GDP) would also have to be included in $\mathbf{L}_{t^\ast}$. We suggest that an analysis that includes all measured variables in $\mathbf{L}_{t^\ast}$ can serve as a useful sensitivity analysis to explore the extent to which effect estimates may change under different assumptions.
\item \textbf{Economic theory approach} (\textit{EconDAG}): A further approach, termed \textit{EconDAG}, includes only variables that are measured during a particular 2-yearly transmission cycle, as defined by our DAG. That is, for the Q-model at $t^{\ast}$, every measured variable between $t^{\ast}-2$ and $t^{\ast}-1$ is included, while for the estimation of the g-model at $t^{\ast}-2$, only variables during the respective cycle are considered. As above, given the assumed time ordering, only variables from the past, and not from the future, are utilized in the respective models.

\end{enumerate}

Given the complexity of the data-generating process, it makes sense to use machine learning techniques to estimate the respective g- and Q-models. For a specified set of learning algorithms and a given set of data, the method minimizing the expected prediction error (as estimated by $k$-fold cross validation) could be chosen. As the best algorithm in terms of prediction error may depend on the given data set, it is often recommended to use super learning instead -- and this is what we use for i), ii) and iii). Super learning \citep{van2007super} (or ``stacking'', \cite{Breiman:1996}) considers a set of learners; instead of picking the learner with the smallest prediction error, one chooses the convex combination of learners that minimizes the $k$-fold cross validation error (for a given loss function, we use $k=10$). The weights relating to this convex combination can be obtained with non-negative least squares estimation (which is implemented in the $R$-package \texttt{SuperLearner}, \cite{Polley:2017}). It can be shown that this weighted combination will perform asymptotically at least as well as the best algorithm, if not better, given that no correctly specified parametric model is contained in the set of learners (\citealp{vanderLaan:2008}).

As described in Section \ref{sec:methods_data_adaptive}, the challenge of model specification, including the choice of appropriate learners and screening algorithms, is to address the complex nonlinear relationships in the data and the $p>n$ problem.

Our strategy is to use the following algorithms: the arithmetic mean of the outcome; generalized linear models (with main terms only and including all two-way interactions); Bayesian generalized linear models with an independent Gaussian prior distribution for the coefficients; classification and regression trees; multivariate adaptive (polynomial) regression splines; generalized additive models; Breimans' random forest; generalized boosted regression modeling; and single-hidden-layer neural networks. The algorithms are carefully chosen to reflect a balance between simple and computationally efficient strategies and more sophisticated approaches that are able to model highly nonlinear relationships and higher-order interactions that may be prevalent in the data. Furthermore, parametric, semiparametric and nonparametric approaches were applied to allow for enough flexibility with respect to committing to parametric assumptions. In particular, tree-based procedures were chosen to handle challenges that frequently come with economic data -- for instance outliers. In addition, since some of the continuous predictors are transformed by the natural logarithm, this strict monotone transformation may affect its variable importance in a regression-based procedure, while trees are not impaired in that respect.

For strategies i)-iii), we use the following learning and screening algorithms:
\begin{enumerate}[a)]
\item \textbf{Screening algorithms:} Used only for estimation approach ii) because of the large covariate set compared to the sample size; we used the elastic net \citep{elasticNet}, the random forest \citep{randomForest}, Cramer's V (with either 4 or 8 variables selected at a maximum) and the Pearson correlation coefficient. The screening algorithms were chosen such that at least a subset of them could handle both categorical and quasi-continuous variables well.
\item \textbf{Learning algorithms:} The 11 learning algorithms mentioned above are the same for estimation strategies i) and iii). i) and iii) were thus estimated with 11 algorithms each. In contrast, strategy ii) additionally benefited from the 5 screening algorithms mentioned in a) where each screening algorithms was run prior to each learning algorithm. We omitted generalized boosted regression modeling from the learner set such that $50 = 5 \times (11 - 1)$ algorithms (i.e.  $\{\text{Screener},\text{Learner}\}$ tuples) emerged. In addition, learning algorithms that are applicable in the $p>n$ case were added without prior screening to the 50 tuples. As a result, when Breimans’ random forest and single-hidden-layer neural networks were added without screening, 52 algorithms could be used for strategy ii); see also Figure \ref{fig:WeightsScreenLearn} in the Appendix.
\end{enumerate}

All estimates have been obtained using the \texttt{ltmle} package in $R$ (\citealp{LTMLEPackage}).

\subsection{Results}\label{sec:data_results}
Descriptive summaries of the data are given in the appendix, in Figures \ref{fig:CBI_regime_switches}-\ref{fig:DynmTreatmentSupp}. They show the variables' distribution over time. Between 1998 and 2010 most measured variables show interesting patterns and changes. For example, one can see a continuously aging population in the countries included, as well as increased levels of central bank transparency. There is support in the data for all three treatment strategies, with 23 countries having an independent central bank throughout the whole time period, 27 countries never having an independent central bank for the period considered and 16 countries which experienced periods with a negative median inflation rate or median inflation above 5\% in the last seven years during 1998 and 2010, while having legislated an independent central bank during the same time period (Figure \ref{fig:DynmTreatmentSupp}).

A naive analysis comparing the mean reductions in inflation between 2000 and 2010 between those countries that had an independent central bank (from 1998 to 2008) and those that had a dependent central bank led to the following results: the mean reduction was 2.3 percentage points for those with an independent central bank, compared to 1.0 percentage points for those with a dependent central bank. This equates to a difference of 1.3 percentage points (95\% CI: -6.1; 3.5). However, such a crude comparison does not allow a causal interpretation and is not an estimate of $\psi_{1,3}$.

\begin{figure}[H]
    \centering
    %Results at first JCI submission:
    %\includegraphics[width=\textwidth]{ATE_infl.pdf}
    \includegraphics[width=\textwidth]{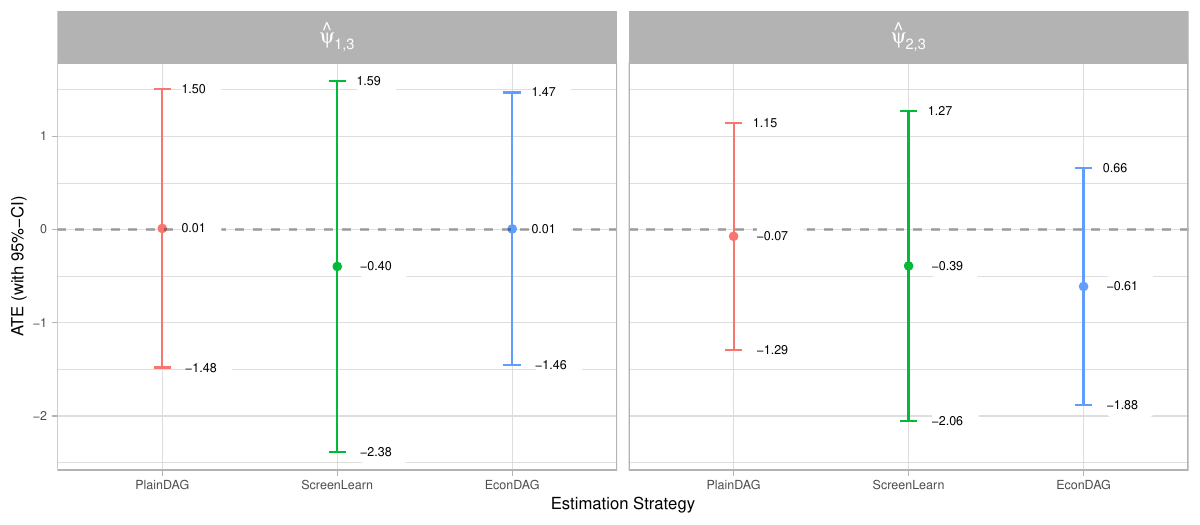}
\caption[]{$\hat{\psi}_{1,3}$ and $\hat{\psi}_{2,3}$ for the three different estimation strategies}
    \label{fig:ATE1}
\end{figure}

The results of the analyses described in Section \ref{sec:data_estimation_LTMLE} are visualized in Figure \ref{fig:ATE1}.

Our main analysis (\textit{PlainDAG}) suggests that if a country had legislated CBI for every year between 1998 and 2008, it would have had an average increase in inflation of 0.01 (95\% confidence interval (CI): -1.48; 1.50) percentage points in 2010. The other two approaches led to slightly different results: -0.44 (95\% CI: -2.38; 1.59) for \textit{ScreenLearn} and 0.01 (95\% CI: -1.46; 1.47) for \textit{EconDAG}.

Similarly, when considering the estimation strategy \textit{PlainDAG}, we can conclude that if a country had legislated an independent central bank for every year when the median of the past seven years of inflation had been above 5\% or below 0\% from 1998 to 2008, it would have achieved an average reduction in inflation of 0.07 percentage points (95\% CI: -1.29; 1.15) in 2010 compared to a central bank that was independent during the same time span (that is, dichotomized CBI = 0). The other two strategies suggest somewhat stronger inflation reductions.

Our findings can be summarized as follows: First, depending on the degree of structural assumptions imposed, we find that an independent central bank has either a negative or no effect on inflation. Second, as suggested by the confidence intervals, we cannot exclude the possibility of a strong negative or positive ATE. Third, the largest estimated ATE (in absolute terms) amounts to -0.61 percentage points (EconDAG). From a monetary policy perspective this can be considered as substantial, given that our study period covers an era of overall low to moderate inflation (characterized by a median inflation rate of about 4\%).

For a sensitivity analysis, we stratified our sample according to the World Bank's income classification into high income (n = 26) and low income (n = 34) countries and reran all analyses. The results are reported in the appendix (cf. Figure \ref{fig:ATE_highIncome} and \ref{fig:ATE_lowIncome}). For high-income countries, the ATE (averaged across estimation strategies) is slightly positive. In contrast, for low-income countries, where inflation has typically been higher, we obtain almost no effect for the static treatment strategy (i.e. $\hat{\Psi}_{1,3}$) and an average reduction of about -0.4 percentage points for the dynamic treatment strategy $\hat{\Psi}_{2,3}$. However, due to the small sample sizes, these results need to be interpreted with caution.

The diagnostics for all analyses are given in Table \ref{tab:CCSummaryStatistics} and Figure \ref{fig:Cum_g_densities} in the Appendix. The cumulative product of inverse probabilities was never below the truncation level of $0.01$, which was re-assuring (\citealp{Petersen:2012}). The maximum value of clever covariates, as defined in (\ref{eqn:clever_covariate}), was always well below 5, which suggests that the chosen super learning approach worked well. However, the mean clever covariate, which is supposed to be broadly approximately 1, was not ideal for dynamic treatment strategy 2, suggesting that $\psi_{2,3}$ should be interpreted cautiously.

%%%%% Distribution of the Clever Covariate %%%%%

\begin{table}[H]
\centering
\begin{adjustbox}{max width=\textwidth}
\begin{tabular}{l|cc|cc|cc|cc|cc|cc|}
\cline{2-13}
  &\multicolumn{2}{c|}{ScreenLearn, $\hat{\psi}_{1,3}$} & \multicolumn{2}{c|}{ScreenLearn, $\hat{\psi}_{2,3}$} & \multicolumn{2}{c|}{EconDAG, $\hat{\psi}_{1,3}$} & \multicolumn{2}{c|}{EconDAG, $\hat{\psi}_{2,3}$} &  \multicolumn{2}{c|}{PlainDAG, $\hat{\psi}_{1,3}$} &  \multicolumn{2}{c|}{PlainDAG, $\hat{\psi}_{2,3}$} \\
  \hline
Intervention & $\bar{A}_{t^{\ast}} =\bar{d}_{t^{\ast}}^3$ & $\bar{A}_{t^{\ast}}=\bar{d}_{t^{\ast}}^1$ & $\bar{A}_{t^{\ast}}=\bar{d}_{t^{\ast}}^3$ & $\bar{A}_{t^{\ast}}=\bar{d}_{t^{\ast}}^2$ & $\bar{A}_{t^{\ast}}=\bar{d}_{t^{\ast}}^3$ & $\bar{A}_{t^{\ast}}=\bar{d}_{t^{\ast}}^1$ & $\bar{A}_{t^{\ast}}=\bar{d}_{t^{\ast}}^3$ & $\bar{A}_{t^{\ast}}=\bar{d}_{t^{\ast}}^2$ & $\bar{A}_{t^{\ast}}=\bar{d}_{t^{\ast}}^3$ & $\bar{A}_{t^{\ast}}=\bar{d}_{t^{\ast}}^1$ & $\bar{A}_{t^{\ast}}=\bar{d}_{t^{\ast}}^3$ & $\bar{A}_{t^{\ast}}=\bar{d}_{t^{\ast}}^2$  \\
\toprule

%% Results for the first JCI submission
%  \hline
%\rowcolor{Gray}
%  Trunc. (\%) & 0.000 & 0.000 & 0.000 & 0.000 & 0.000 & 0.000 & 0.000 & 0.000 & 0.000 & 0.000 & 0.000 & 0.000 \\
%  CC Mean & 0.823 & 0.760 & 0.803 & 0.457 & 0.846 & 0.828 & 0.826 & 0.454 & 0.830 & 0.718 & 0.870 & 0.490 \\
%    \rowcolor{Gray}
%  CC Max.& 3.038 & 3.635 & 3.071 & 1.996 & 3.603 & 4.709 & 3.323 & 2.089 & 3.214 & 3.568 & 3.417 & 2.429 \\
%  \hline
%    CC Mean Max. & 0.939 & 0.983 & 0.879 & 0.502 & 0.874 & 0.942 & 0.914 & 0.487 & 0.878 & 0.863 & 0.924 & 0.553 \\
%  \rowcolor{Gray}
%  CC Mean Min. & 0.735 & 0.602 & 0.731 & 0.432 & 0.763 & 0.642 & 0.749 & 0.405 & 0.781 & 0.621 & 0.780 & 0.442 \\
\hline
\rowcolor{Gray}
Trunc. (\%) & 0.00 & 0.00 & 0.00 & 0.00 & 0.00 & 0.00 & 0.00 & 0.00 & 0.00 & 0.00 & 0.00 & 0.00 \\
  CC Mean & 0.89 & 0.90 & 0.91 & 0.52 & 0.83 & 0.71 & 0.83 & 0.50 & 0.82 & 0.72 & 0.82 & 0.50 \\
      \rowcolor{Gray}
  CC Max. & 3.64 & 4.88 & 3.71 & 2.57 & 2.27 & 2.63 & 2.30 & 1.97 & 2.25 & 2.63 & 2.26 & 1.99 \\
  CC Mean Max. & 0.94 & 1.02 & 0.95 & 0.59 & 0.85 & 0.77 & 0.86 & 0.51 & 0.85 & 0.75 & 0.86 & 0.51 \\
      \rowcolor{Gray}
  CC Mean Min. & 0.81 & 0.77 & 0.88 & 0.48 & 0.79 & 0.62 & 0.79 & 0.47 & 0.78 & 0.64 & 0.79 & 0.47 \\
\bottomrule
\end{tabular}
    \end{adjustbox}
  \caption{Row 1: mean percentage of observations that had to be truncated because the cumulative product of inverse probabilities was $<0.01$. Rows 2 and 3: Mean and maximum value of the clever covariate. All results are averaged over the 5 imputed data sets. Rows 4 and 5 contain the minimum and maximum of the five mean clever covariate values across the imputed data sets.}
  \label{tab:CCSummaryStatistics}
\end{table}

Figure \ref{fig:WeightsScreenLearn} (Appendix) visualizes the learner weight distribution. In our analysis, a multitude of learners and screening algorithms were important, including neural networks, random forests, regression trees and Bayesian generalized linear models.

\section{Simulations}\label{sec:simulations}

Motivated by our data analysis, we explore the extent to which model misspecification and choice of learner sets may affect effect estimation with longitudinal maximum likelihood estimation (and competing methods).

\subsection{Data-Generating Processes}\label{sec:DGPs}

We specified two data-generating processes: a simple one with 3 time points and one time-dependent confounder and a more complex one with up to 6 time points and 10 time-varying variables.

For the first simulation (\textit{Simulation 1}), we assume the following time ordering:
\begin{eqnarray*}
     O = (L_{1},A_1,Y_1,L_2,A_2,Y_2,L_3,A_3,Y_3)
\end{eqnarray*}

Using the $R$-package \texttt{simcausal} (\citealp{Sofrygin:2016}), we define preintervention distributions as listed in Table \ref{tab:DGPSimulation1} (Appendix).

For the second simulation (\textit{Simulation 2}), we use the following time ordering:
\begin{eqnarray*}
     O = (L_{1}^1,A_1,Y_1,L_1^2,\ldots,L_1^{10},\ldots,L_{5}^1,A_5,Y_5,L_5^2,\ldots,L_5^{10},L^1_6,A_6,Y_6)
\end{eqnarray*}

We generated the preintervention data according to the distributions specified in Table \ref{tab:DGPSimulation2} (Appendix).

\subsection{Target Parameter and Interventions}

For both simulations, we were interested in evaluating ATEs between two static interventions. That is, we were interested in

\vspace*{-0.25cm}
\begin{flalign*}
\bar{d}_{t^{+}}^{Sim1,1} \phantom{\quad\,\,\bar{L}^1_{t^{\ast}-1}} &= \,\, \left\{a_{t^{+}}=1 \quad \forall t^{+} \in \{1,2,3\}\right.&
\end{flalign*}\vspace*{-1.25cm}
\begin{flalign*}
\bar{d}_{t^{+}}^{Sim1,0} \phantom{\,\,\quad\bar{L}^1_{t^{\ast}-1}} &= \,\, \left\{a_{t^{+}}=0 \quad \forall t^{+} \in \{1,2,3\}\right.&
\end{flalign*}
and
\vspace*{-0.15cm}
\begin{flalign*}
\bar{d}_{t^{++}}^{Sim2,1} \phantom{\quad\,\,\bar{L}^1_{t^{\ast}-1}} &= \,\, \left\{a_{t^{++}}=1 \quad \forall t^{++} \in \{1,2,3,4,5,6\}\right.&
\end{flalign*}\vspace*{-1.25cm}
\begin{flalign*}
\bar{d}_{t^{++}}^{Sim2,0} \phantom{\,\,\quad\bar{L}^1_{t^{\ast}-1}} &= \,\, \left\{a_{t^{++}}=0  \quad \forall t^{++} \in \{1,2,3,4,5,6\}\right.&
\end{flalign*}

The target parameters of interest are thus
\begin{eqnarray}
\psi_{1} = \mathbb{E}(Y_{2}^{\bar{d}_{t+}^{Sim1,1}})-\mathbb{E}(Y_{2}^{\bar{d}_{t+}^{Sim1,0}})\,,\quad
\psi_{2} = \mathbb{E}(Y_{6}^{\bar{d}_{t++}^{Sim2,1}})-\mathbb{E}(Y_{6}^{\bar{d}_{t++}^{Sim2,0}}) \,,
\end{eqnarray}

\subsection{Estimations}

In our primary analysis, we used longitudinal targeted maximum likelihood estimation for both simulations. In a secondary analysis, we also evaluated the performance of (longitudinal) inverse probability of treatment weighting (see, e.g., \citealp{Daniel:2013} and the references therein).

For LTMLE, we considered four different estimation approaches, the first for the first simulation and another three for the second simulation:

\begin{enumerate}[i)]
\item Estimation as explained in Section \ref{sec:methods_estimation_LTMLE}. Q- and g-models were fitted with (generalized) linear models. This is estimation approach \textit{GLM}.
\item Estimation as explained in Section \ref{sec:methods_estimation_LTMLE}. Q- and g-models were fitted with a data-adaptive approach using super learning. There were four candidate learners: the arithmetic mean, GLMs, Bayesian generalized linear models with an independent Gaussian prior distribution for the coefficients, as well as classification and regression trees. No screening of variables was conducted. This is estimation approach \textit{L1}.
\item Estimation as explained in Section \ref{sec:methods_estimation_LTMLE}. Q- and g-models were fitted with a data-adaptive approach using super learning. The same four learners as in \textit{L1} are utilized; however, variable screening with Pearson's correlation coefficient was conducted. In addition, four more learners were added: multivariate adaptive (polynomial) regression splines \citep{friedman1991MARS}, generalized additive models, and generalized linear models including the main effects with all corresponding two-way interactions. These additional four learners included variable screening with the elastic net ($\alpha = 0.75$). This is estimation approach \textit{L2}.
\item Estimation as explained in Section \ref{sec:methods_estimation_LTMLE}. Q- and g-models were fitted with a data-adaptive approach using super learning. The eight learning/screening combinations from \textit{L2} were used. In addition, single-hidden-layer neural networks were used, once without variable screening and once with elastic net screening. Finally, the last learner is composed of classification and regression with the random forest. This is estimation approach \textit{L3}.
\end{enumerate}

We also obtained estimates for the ATE based on IPTW. The estimation of the propensity scores was identical to the estimation of the g-models within LTMLE and is thus also based on the estimation procedures described in i)-iv).

\subsection{Comparisons}
We compared the estimated absolute (abs.) bias and coverage probabilities for the estimated ATEs for the two simulations and for both correctly and incorrectly specified Q-models (see details below).

\begin{enumerate}[i)]
\item \textbf{Simulation 1:} The incorrect, misspecified, Q-models omit $\mathbf{L}:=(L_1, L_2, L_3)$ entirely. By contrast, the g-models were specified such that the entire covariate histories are taken into account. As a result, if no screening is applied (estimation strategies GLM and L1), all relevant variables are used for estimation; however, with screening (estimation strategies L2 and L3), some variables might be omitted.

\item \textbf{Simulation 2:} The incorrect, misspecified, Q-models do not use $\mathbf{L^1} := (L^1_1, L^1_2, L^1_3, L^1_4, L^1_5, L^1_6, L^1_7)$ for estimation. Thus, one relevant back-door path remains unblocked, which leads to time-dependent confounding with treatment-confounder feedback. As in simulation 1, all g-models were specified such that the entire covariate histories are taken into account.
\end{enumerate}

\subsection{Results}

\begin{figure}[H]
    \centering
    \includegraphics[width=\textwidth]{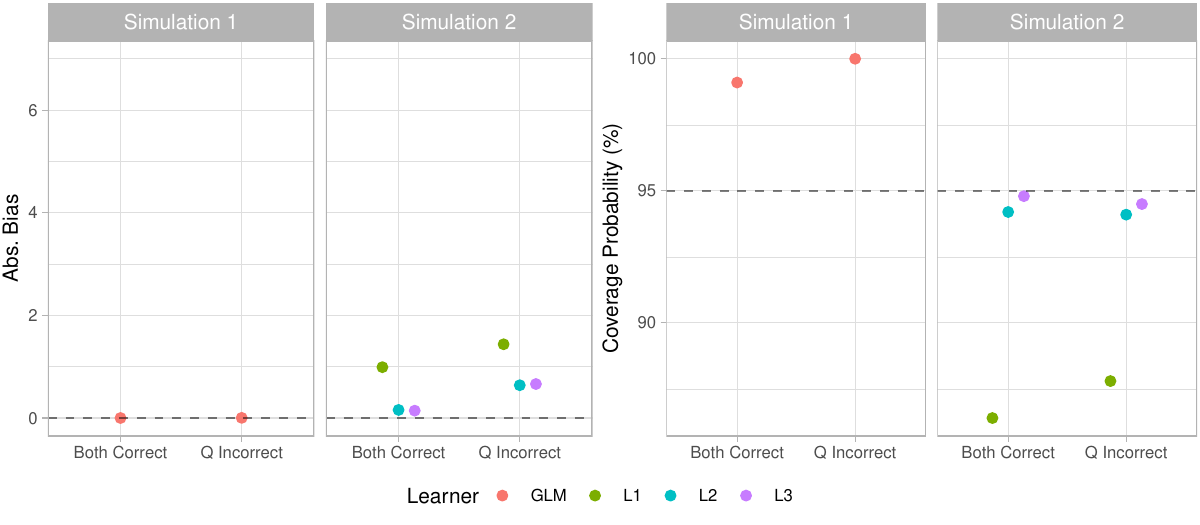}
\caption[]{Absolute bias and coverage probability for both simulations -- for correctly specified Q- and g-models (\textit{Both Correct}) and misspecified Q-models (\textit{Q Incorrect}) of LTMLE.}
    \label{fig:SimBiasCP}
\end{figure}

The results after 1000 simulation runs are summarized in Figure \ref{fig:SimBiasCP}.

In simulation 1, LTMLE provides approximately unbiased estimates even under misspecified Q-models. This is because targeted maximum likelihood estimation is a doubly robust estimator, and thus misspecification of either the Q- or g-models can be handled. However, the coverage probabilities are too high. See \cite{Tran:2018} for a discussion of this issue.

Under the more complex setup of simulation 2, there is small bias if both the Q- and g-models contain the relevant adjustment variables (\textit{Both Correct}) and learner set \textit{L1} is used (Bias = 0.991). The more sophisticated learner sets \textit{L2} and \textit{L3} yield much better estimates (Bias = 0.158 and 0.144). With incorrect specification of the Q-model, there is again some bias (Bias = 1.438, 0.639, 0.663). Interestingly, for simulation 2, the most complex estimation approach with the largest learner set \textit{L3} does not produce a substantial improvement over \textit{L2}. This highlights that a simple increase in learners does not necessarily improve the finite sample performance of LTMLE, although sufficient breadth and complexity is certainly always needed, as seen by the inferior performance of the first learner set.

In simulation 1, the confidence intervals have too large coverage probabilities. However, in simulation 2, using \textit{L2} and \textit{L3} yields (close to) nominal coverage probabilities. Nevertheless, our results highlight the need to develop more reliable variance estimators, such that overall better coverage can be achieved.

Note that while LTMLE may produce approximately unbiased point estimates, IPTW does not seem to benefit from complex estimation procedures for the propensity scores (g-models) in the second simulation. The estimates are rather volatile, with some bias and poor coverage probabilities. These conclusions hold for all learner sets considered (Appendix, Figure \ref{fig:SimBiasCP_IPTW}).

\section{Conclusions}\label{sec:conclusions}
We have shown that even for complex macroeconomic questions, it is possible to develop a causal model and implement modern doubly robust longitudinal effect estimators. We believe that this is an important contribution in light of the current debate on the appropriate implementation and use of causal inference for economic questions (\citealp{Imbens:2019}). Our suggestion was to commit to a causal model, motivate it in substantial detail (as in Appendix \ref{sec:appendix_justification}), discuss possible violations of it, and ultimately conduct sensitivity analyses that evaluate effect estimates under different (structural) assumptions.

While the statistical literature has emphasized the benefits of doubly robust effect estimation in conjunction with extensive machine learning (\citealp{vanderLaan:2011}), its use in sophisticated longitudinal settings has sometimes been limited due to computational challenges and constraints (\citealp{Schomaker:2019}). We have shown how the use of screening and learning algorithms that are tailored to the question of interest can help to facilitate a successful implementation of this approach.

As stressed by \citet{Imbens:2019}: \textit{``[...] models in econometric papers are often developed with the idea that they are useful on settings beyond the specific application in the paper''}. We hope that both our causal model, i.e., the DAG, and our proposed estimation techniques will be useful in applications other than ours.

Our simulation studies suggest that LTMLE with super learning can yield good point estimates compared to competing approaches, even under model misspecification. However, both the coverage of confidence intervals and the appropriate choice of learners are challenges that warrant more investigation. Recent research confirms that the development of more robust variance estimators is urgently needed (\citealp{Tran:2018}) and that learner selection is becoming more diverse (\citealp{Gehringer:2018}).

%From a monetary policy point of view, we conclude that there is no strong support for the hypothesis that an independent central bank necessarily lowers inflation, although our confidence intervals were wide. Future research may investigate whether this finding holds for subgroups of particular countries, such as developing countries, and for different time periods. However, even if the impact of CBI on inflation seems to be weak, independent central banks could still have beneficial effects on outcomes other than those investigated by us.

From a monetary policy point of view, we conclude that based on the estimates from the main analysis there is no strong support for the hypothesis that an independent central bank necessarily affects inflation, although our confidence intervals were wide. Making fewer or different structural assumptions, as in our secondary analyses, leads to an average inflation reduction of up to 0.6 percentage points under central bank independence. An ATE of -0.6 percentage points may be seen as substantial, considering that the period on which our estimations are based is overall characterized by low to moderate inflation. However, a naive use of super learning (as in our ``ScreenLearn'' secondary analysis) may be potentially dangerous because important collider and mediator structures may be overlooked, which can yield different, possibly incorrect results. A comparison of the point estimates from the main and secondary analysis reflects this consideration. As highlighted throughout this paper, while a sophisticated computational approach can be advantageous for doubly robust causal effect estimation, it can not replace the commitment to well-thought-out structural assumptions about the macroeconomic process under consideration.

% Literature
\addcontentsline{toc}{section}{References}

\bibliographystyle{agsm-mod}%{plainnat}
{\footnotesize{
\bibliography{bibliography}
}}

\clearpage
\appendix

\newgeometry{
  left=0.5cm,
  right=0.5cm,
  top=0.5cm,
  bottom=1.5cm
}

\section{More details on the causal model}\label{sec:appendix_arrows_DAG}
\subsection{Definition of the variables listed in the DAG}

{\scriptsize{
\input{VertexReasoning}
}}

\clearpage

\subsection{Explanation for the arrows in the DAG}\label{sec:appendix_justification}
{\scriptsize{
 \input{EdgeReasoning}
}}

\clearpage

\section{Additional Material related to the Data Analysis}\label{sec:appendix_data}

\fontsize{8}{8}\selectfont
\begin{figure}[H]
    \centering
    \includegraphics[width=\textwidth]{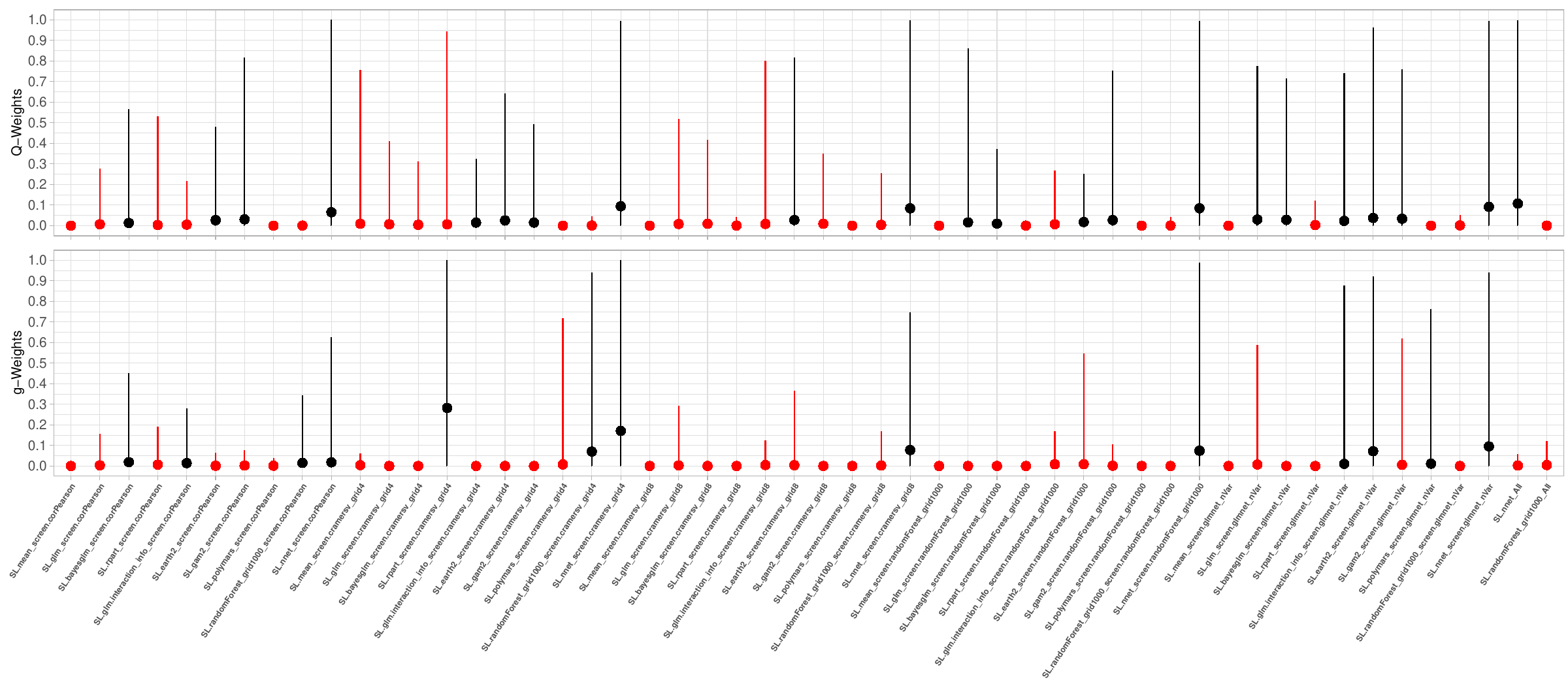}
    \caption[]{Distribution of learner weights. The visualized distributions are based on the merged learner weights that resulted from the estimation of $\Psi_{1,3}$ and $\Psi_{2,3}$ ($\bar{d}_{t^{\ast}}^{1} , \bar{d}_{t^{\ast}}^{2}$ and twice $\bar{d}_{t^{\ast}}^{3}$), summarized across the imputed data sets. The plotted point represents the mean of each distribution. If it is below 0.01, both the distribution and the mean are displayed in red.}
    \label{fig:WeightsScreenLearn}
\end{figure}

\begin{figure}[H]
    \centering
    \includegraphics[width=\textwidth]{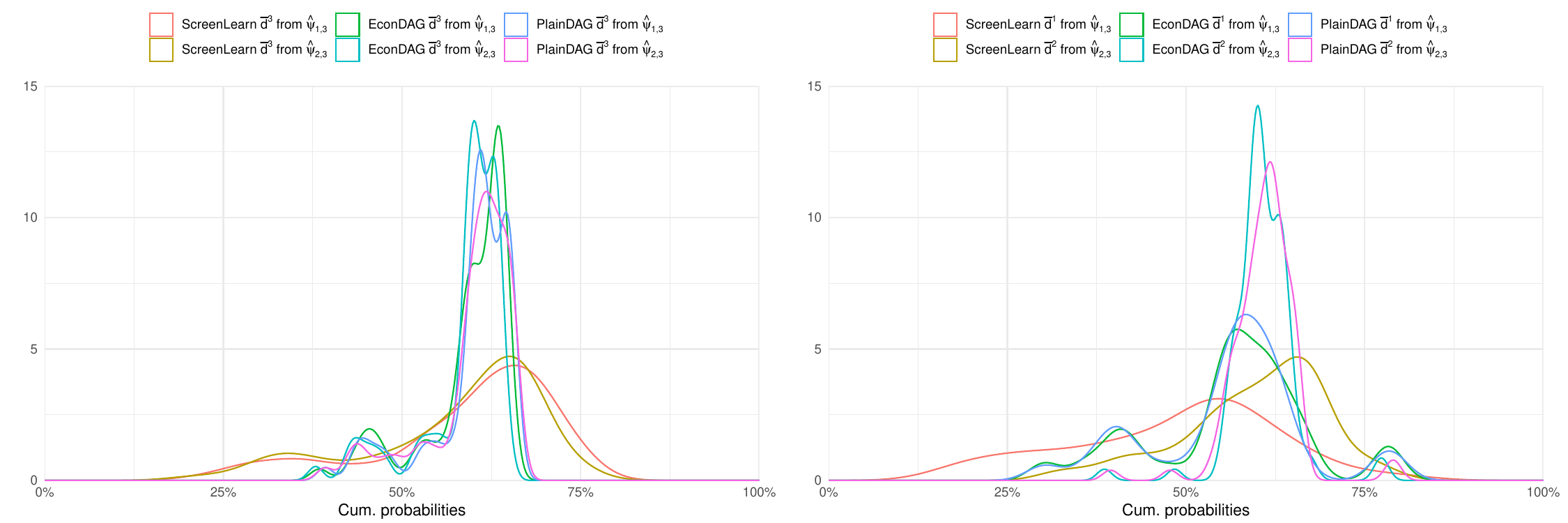}
    \caption[]{Kernel density plots of cumulative treatment probabilities for
    $T = 2010$. In the left panel, estimated probabilities of $\bar{d}_{t^{\ast}}^{3}$ are shown while the right panel shows estimated probabilities for $\bar{d}_{t^{\ast}}^{1}$ and $\bar{d}_{t^{\ast}}^{2}$ are depicted.}
    \label{fig:Cum_g_densities}
\end{figure}

\begin{figure}[H]
    \centering
    \includegraphics[width=\textwidth]{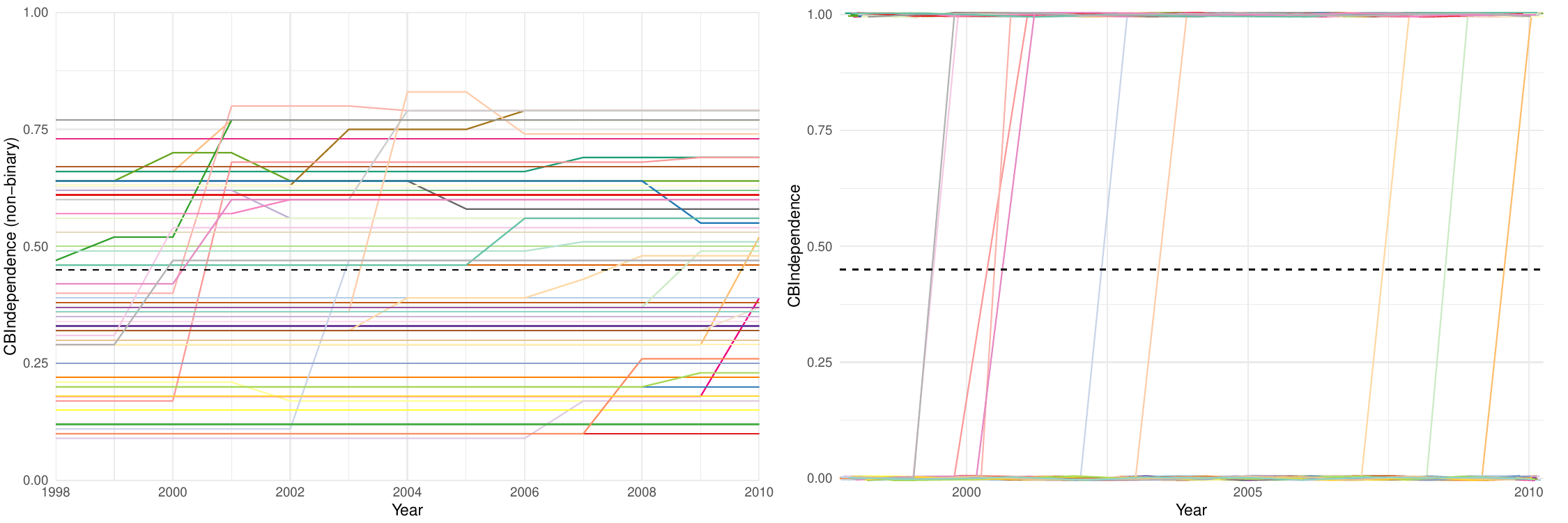}
    \caption[]{Trajectories of the intervention variable (CBIndependence) for all included countries ($n=60$).}
    \label{fig:CBI_regime_switches}
\end{figure}

\begin{figure}[H]
    \centering
    \includegraphics[width=\textwidth]{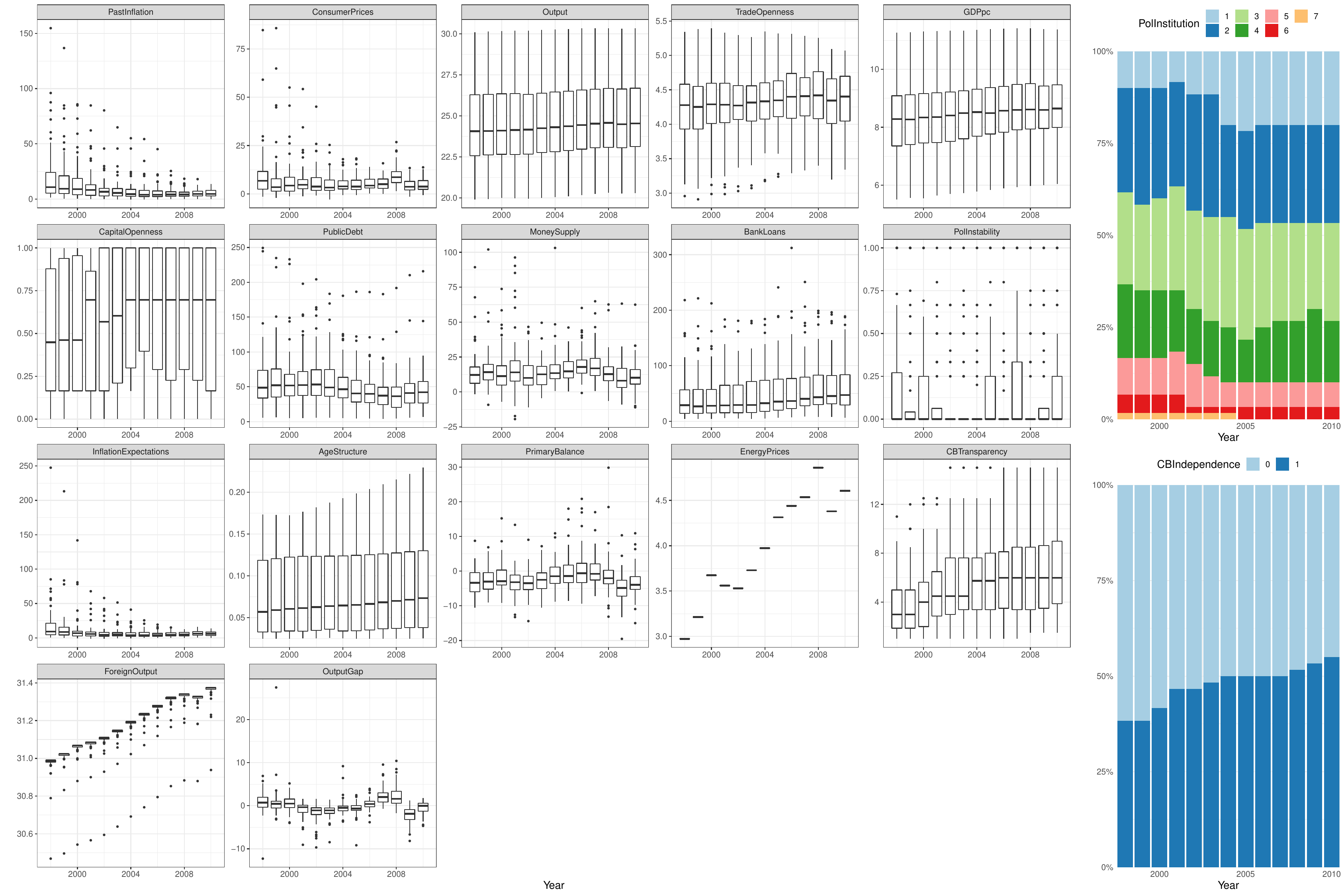}
    \caption[]{Summary statistics for all variables included in the data analysis. The variables "Output", "TradeOpenness", "GDPpc", "EnergyPrices" and "ForeignOutput" were transformed by the natural logarithm for better readability. Appendix \ref{sec:appendix_arrows_DAG} gives more details on the variables and what they measure.}
    \label{fig:DescrVars}
\end{figure}

\begin{figure}[H]
    \centering
    \includegraphics[width=\textwidth]{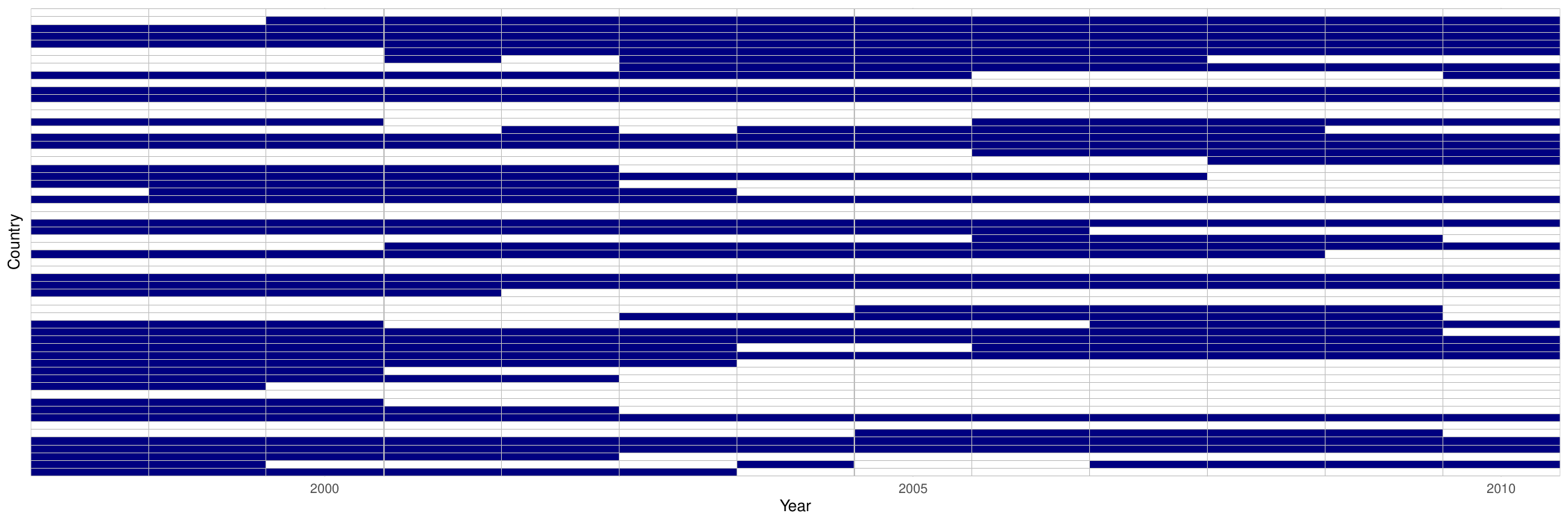}
    \caption[]{Blue tiles indicate when a country has had a negative or above 5\% median inflation rate in the last seven years while having legislated an independent central bank simultaneously.}
    \label{fig:DynmTreatmentSupp}
\end{figure}

\begin{figure}[H]
    \centering
    \includegraphics[width=\textwidth]{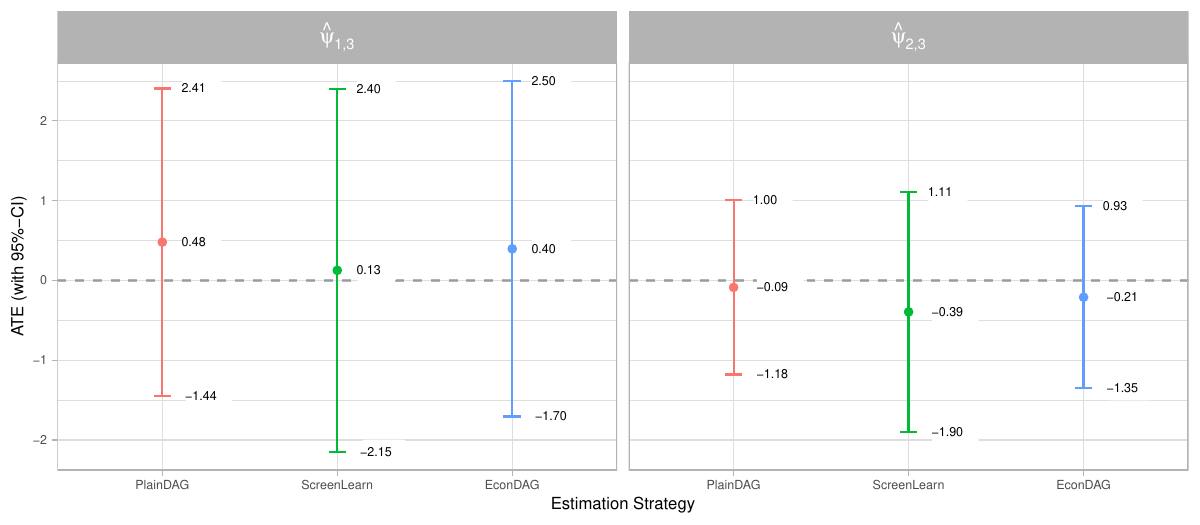}
    \caption[]{ATE among the high-income countries (n = 26).}
    \label{fig:ATE_highIncome}
\end{figure}

\begin{figure}[H]
    \centering
    \includegraphics[width=\textwidth]{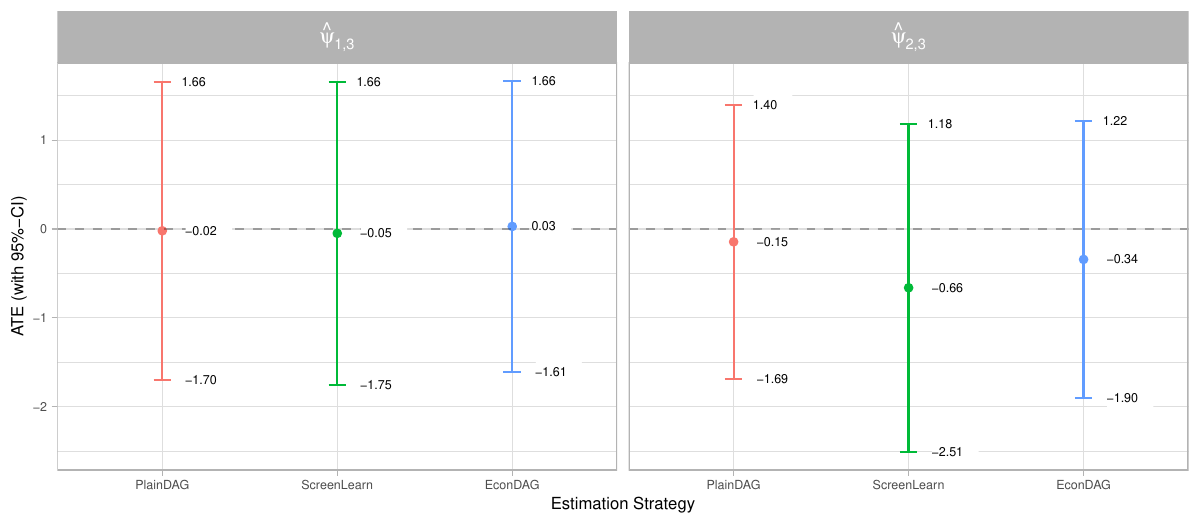}
    \caption[]{ATE among the low-income countries (n = 34).}
    \label{fig:ATE_lowIncome}
\end{figure}

%%%%% Distribution learner weights for the case of inflation %%%%%
%\begin{figure}[H]
%    \centering
%    \includegraphics[width=\textwidth]{EstimationPlots/PlainDAGWeights.pdf}
%    \caption[]{PlainDAG: Learner Weights of Q and g Estimation (Treatment and Control) for merged $\Psi_{1,3}$ and $\Psi_{2,3}$}
%    \label{fig:WeightsPlainDAG}
%\end{figure}

%\begin{figure}[H]
%    \centering
%    \includegraphics[width=\textwidth]{EstimationPlots/EconDAGWeights.pdf}
%    \caption[]{EconDAG: Learner Weights of Q and g Estimation (Treatment and Control) for merged $\Psi_{1,3}$ and $\Psi_{2,3}$}
%    \label{fig:WeightsEconDAG}
%\end{figure}

\section{Details on the Simulation Study}\label{sec:appendix_simulation}

\subsection{IPTW}\label{sec:Learner_Weights}

\begin{figure}[H]
    \centering
    \includegraphics[width=\textwidth]{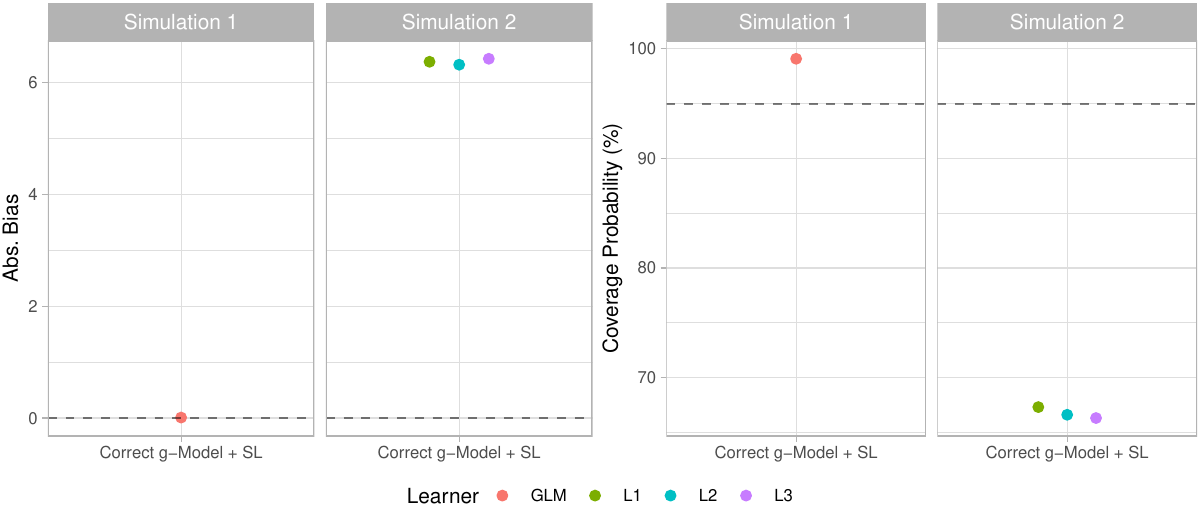}
    \caption[]{Absolute bias and coverage probabilities for estimation with IPTW. Bias: 0.009 (GLM), 6.377 (L1), 6.325 (L2), 6.431 (L3) and coverage probability: 99.1 \% (GLM), 67.3 \% (L1), 66.6 \% (L2), 66.3 \% (L3).}
    \label{fig:SimBiasCP_IPTW}
\end{figure}

\subsection{Data-Generating Processes (DGP)}\label{sec:DGP}

\fbox{
{\small{
\begin{tabular}{p{0.4\textwidth}|p{0.5\textwidth}}
$t=1$ & $t=2,3$ \\
\hline
$L_{t} \sim N(0,0.25)$ &  $L_{t} \sim N(L_{t-1}+A_{t-1},0.25)$ \\
$A_{t} \sim B(\text{expit}(L_{t}))$ &  $A_{t} \sim B(\text{expit}(L_{t}+ 2 \times A_{t-1} -L_{t-1}))$ \\
$Y_{t} \sim N(50\times A_{t} + L_{t},0.25)$ &  $Y_{t} \sim N(50\times A_{t} + L_{t} + L_{t-1} + Y_{t-1},0.06)$ \\
\end{tabular}
}}}
\captionof{table}{DGP for Simulation 1}
\label{tab:DGPSimulation1}

\fbox{
{\small{
\begin{tabular}{p{0.4\textwidth}|p{0.5\textwidth}}
$t=1$ & $t=2,\dots,6$ \\
\hline
$L^{1}_{t} \sim N(0,0.25)$ & $L^{1}_{t} \sim N(L^{7}_{t-1},0.25)$ \\
$A_{t} \sim B(\text{expit}(L^{1}_{t}))$ & $A_{t} \sim B(\text{expit}(0.25 \times L^{1}_{t} + 0.25 \times L^{6}_{t-1}))$ \\
$Y_{t} \sim N(A_{t} + L^{1}_{t},9)$ & $Y_{t} \sim N(A_{t} +  L^{1}_{t} + L^{9}_{t-1} + 0.05 \times L^{10}_{t-1},0.25)$ \\
$L^{5}_{t} \sim N(Y_{t},2.25)$ &  \\
& \\
$t=1,\dots,5$ & $t=1,\dots,5$ \\
\hline
$L^{2}_{t} \sim N(A_{t} + L^{1}_{t},0.25)$ & $L^{5}_{t} \sim N(Y_{t} + L^{10}_{t-1},2.25)$  \\
$L^{3}_{t} \sim N(Y_{t} + L^{2}_{t},1)$ &  \\
$L^{4}_{t} \sim N(A_{t},0.25)$ &  \\
$L^{6}_{t} \sim N(L^{4}_{t},0.25)$ &  \\
$L^{7}_{t} \sim N(L^{2}_{t},0.25)$ &  \\
$L^{8}_{t} \sim N(L^{5}_{t},0.25)$ &  \\
$L^{9}_{t} \sim N(L^{3}_{t},1)$ &  \\
$L^{10}_{t} \sim N(L^{8}_{t} + L^{9}_{t},0.25)$ &  \\
\end{tabular}
}}}
\captionof{table}{DGP for Simulation 2}
\label{tab:DGPSimulation2}

\end{document}

%% file: DAG.tex
\newgeometry{
  left=0.2cm,
  right=0.2cm,
  top=0.2cm,
  bottom=0.6cm
}

\begin{landscape} % Turn A4 Page 90 Degrees
\begin{figure}
\scalebox{0.53}{ % Scale whole DAG to A4 suited size
\begin{tikzpicture}
\tikzstyle{VertexStyle} = [shape = ellipse, draw,minimum size = 2pt,inner sep=0pt] % Vertex Style
\SetUpEdge[lw = 1pt, color = black, labelcolor = white] % Edge Style
%\GraphInit[vstyle=Normal]
%\SetGraphUnit{3} % Abstand der Knoten untereinander
%\tikzset{VertexStyle/.append  style={fill}}

% Dashed Box Preperation
\tikzset{bigbox/.style={draw, inner sep=5pt,label={[align=center,shift={(5ex,0ex)}]south:\llap{#1}}}}

%% Knoten (=Vertex)
% Pricing
\SetUpVertex[FillColor=green!20]
\Vertex[Math,L={Consumer\ Prices_{t^\ast}\ (Y_{t^\ast})},x=43,y=0] {Inflation}
\SetUpVertex[FillColor=white]
\Vertex[Math,L={Consumption\ Tax_{t^\ast}},x=43,y=2] {VAT}
\Vertex[Math,L={Pricing\ by\ Companies_{t^\ast}},x=35,y=0] {PC}
\Vertex[Math,L={Price\ Mark-Up_{t^\ast-1}},x=35,y=-2] {Mark Up}
\Vertex[Math,L={Market\ Power_{t^\ast-1}},x=43,y=-2] {Market Power}

\SetUpVertex[FillColor=black!20]
\Vertex[Math,L={Output_{t^\ast-1}\ (L^{5}_{t^\ast-1})},x=35,y=-8] {GDP}
\SetUpVertex[FillColor=white]

% HH and company Decisions
\Vertex[Math,L={Savings_{t^\ast-1}},x=31,y=-15] {Savings}
%\Vertex[Math,L={Savings_{t^\ast-2}},x=22.5,y=-16.8] {SavingsT2}
\SetUpVertex[FillColor=black!20]
\Vertex[Math,L={Age\ Structure_{t^\ast-1}\ (L^{1}_{t^\ast-1})},x=31,y=-18] {Aging}
\SetUpVertex[FillColor=white]
\Vertex[Math,L={Nominal\ Wages_{t^\ast-1}},x=35,y=-16] {Last Nominal Wage}
\Vertex[Math,L={Investments_{t^\ast-1}},x=25,y=-16.5] {Investments}
\Vertex[Math,L={Tobin's\ q_{t^\ast-2}},x=19,y=-16.5] {Tobin}
\Vertex[Math,L={Firms'\ net\ worth_{t^\ast-2}},x=19.2,y=-20] {FirmValue}
\Vertex[Math,L={AS\ \&\ MH_{t^\ast-2}},x=19.1,y=-23] {MarketEfficiency}
\Vertex[Math,L={Firms'\ liquid._{t^\ast-2}},x=30,y=-22.4] {Liquidity}
\Vertex[Math,L={Asset\ Prices_{t^\ast-2}},x=21,y=-18] {Assets}
%\Vertex[Math,L={Asset\ Prices_{t^\ast-1}},x=26.5,y=-17] {AssetsT1}
\Vertex[Math,L={Consumption_{t^\ast-1}},x=35,y=-11] {Consumption}
\Vertex[Math,L={Disposable\ Income_{t^\ast-1}},x=35,y=-14] {Disposable Income}
\Vertex[Math,L={Taxes\ and\ Social\ Securities_{t^\ast-1}},x=39,y=-17] {Last TSS}
\Vertex[Math,L={Wealth_{t^\ast-1}},x=35,y=-12.5] {Wealth}

% Public Finance
\SetUpVertex[FillColor=black!20]
\Vertex[Math,L={Public\ Debt_{t^\ast-2}\ (L^{3}_{t^\ast-2})},x=18,y=-14] {PastDebt}
\Vertex[Math,L={Public\ Debt_{t^\ast-2}\ (L^{3}_{t^\ast-2})},x=7,y=-22] {PastDebt2}
\Vertex[Math,L={Public\ Debt_{t^\ast-1}\ (L^{3}_{t^\ast-1})},x=26.5,y=-12.5] {PublicDebt}
\SetUpVertex[FillColor=white]
\Vertex[Math,L={Debt\ Management_{t^\ast-2}},x=9,y=-24.5] {DebtManagement}
\SetUpVertex[FillColor=black!20]
\Vertex[Math,L={Prim.\ Balance_{t^\ast-1}\ (L^{2}_{t^\ast-1})},x=26.5,y=-10.5] {BudgetBalance}
\SetUpVertex[FillColor=white]
\Vertex[Math,L={Fiscal\ Spending_{t^\ast-1}},x=30,y=-9] {Fiscal Spending}
\Vertex[Math,L={Fiscal\ Revenue_{t^\ast-1}},x=17.5,y=-10.5] {Fiscal Revenue}
\Vertex[Math,L={Taxes\ and\ Social\ Security_{t^\ast-1}},x=17.5,y=-12] {TASS1}
\Vertex[Math,L={Consumption\ Tax_{t^\ast-1}},x=17.5,y=-9] {VAT1}
%\Vertex[Math,L={FP\ Decision_{t^\ast}},x=13,y=-8.5] {Fiscal Policy}

% Globalization
\Vertex[Math,L={Net\ Exports_{t^\ast-1}},x=43,y=-16] {Net Exports}
\SetUpVertex[FillColor=black!20]
\Vertex[Math,L={Foreign\ Output_{t^\ast-1}\ (L^{4}_{t^\ast-1})},x=43,y=-12] {Foreign Output}
\SetUpVertex[FillColor=white]
\Vertex[Math,L={Real\ Exchange\ Rate_{t^\ast-2}},x=43,y=-19] {Past X-Rate}
\SetUpVertex[FillColor=black!20]
\Vertex[Math,L={Trade\ Openness_{t^\ast-2}\ (L^{10}_{t^\ast-2})},x=3,y=-28] {TradeOpenness}
\SetUpVertex[FillColor=white]
\Vertex[Math,L={Share\ of\ Non-Tradables_{t^\ast-2}},x=14,y=-26] {ShareNonTrade}
\SetUpVertex[FillColor=black!20]
\Vertex[Math,L={Past\ Inflation_{t^\ast-2}\ (L^{9}_{t^\ast-2})},x=0,y=-25.5] {MovingAverage}
\Vertex[Math,L={Consumer\ Prices _{t^\ast-2}\ (L^{8}_{t^\ast-2})},x=42,y=-24.5] {Past Prices}
\SetUpVertex[FillColor=white]

% Production Cost
\Vertex[Math,L={Production\ Cost_{t^\ast-1}},x=28,y=0] {Production Cost}
\Vertex[Math,L={Non-Labor\ Costs_{t^\ast-1}},x=28,y=-1.5] {NLC}
\SetUpVertex[FillColor=black!20]
\Vertex[Math,L={Energy\ Prices_{t^\ast-1}\ (L^{7}_{t^\ast-1})},x=18,y=-1.5] {Energy Price}
\SetUpVertex[FillColor=white]
\Vertex[Math,L={Technological\ Progress_{t^\ast-2}},x=28,y=-3] {Machine Productivity}
\Vertex[Math,L={Technological\ Progress_{t^\ast-1}},x=28,y=-4] {Machine Productivity1}
\Vertex[Math,L={Technological\ Progress_{t^\ast, \dots t+8}},x=30.5,y=-5.5] {Machine ProductivityFuture}

% Wages
\Vertex[Math,L={Labor\ Costs_{t^\ast-1}},x=18,y=0] {LC}
\Vertex[Math,L={Taxes\ and\ Social\ Security_{t^\ast-1}},x=18,y=2] {TASS}
\Vertex[Math,L={Nominal\ Wages_{t^\ast-1}},x=9,y=0] {Nominal Wages}
\Vertex[Math,L={Bargaining\ Power_{t^\ast-2}},x=9,y=-1.5] {EBP}
\Vertex[Math,L={Bargaining\ Power_{t^\ast-1}},x=9,y=-2.7] {EBP1}
\Vertex[Math,L={Labor\ Productivity_{t^\ast-1}},x=12.5,y=-4.5] {LProductivity}
\Vertex[Math,L={Labor\ Productivity_{t^\ast, \dots t+8}},x=12.5,y=-8] {LProductivityFuture}
\SetUpVertex[FillColor=black!20]
\Vertex[Math,L={Output\ Gap_{t^\ast-1}\ (L^{6}_{t^\ast-1})},x=18,y=-4] {BC1}
\SetUpVertex[FillColor=white]
\Vertex[Math,L={Labor\ Unions_{t^\ast-1}},x=7,y=-4.5] {LUnion}

\Vertex[Math,L={Human\ and\ Public\ Capital_{t^\ast-1}},x=21,y=-8] {RAD}
\Vertex[Math,L={Human\ and\ Public\ Capital_{t^\ast-10,\dots , t-2}},x=18,y=-6] {RADT2}

% Inflation Expectations
\SetUpVertex[FillColor=black!20]
\Vertex[Math,L={Past\ Inflation_{t^\ast-2}\ (L^{9}_{t^\ast-2})},x=0.5,y=-2] {MovingAverage1}
%\Vertex[Math,L={Past\ Inflation_{t^\ast-2}},x=1,y=-4.5] {PastInflation1}
\Vertex[Math,L={Inflation\ Expectations_{t^\ast-2}\ (L^{17}_{t^\ast-2})},x=4.5,y=-6] {Inflation Expectation}
\SetUpVertex[FillColor=white]
%\Vertex[Math,L={Inflation\ Memory_{t^\ast}},x=2,y=-4] {Inflation Memory}

% Monetary Strategies
\Vertex[Math,L={CB\ Credibility_{t^\ast-2}},x=4.5,y=-10] {CBC}
\SetUpVertex[FillColor=white]
\Vertex[Math,L={Exchange-Rate\ Regime_{t^\ast-2}},x=10.5,y=-17.5] {ERA}
\Vertex[Math,L={Targeting\ Regime_{t^\ast-2}},x=4.5,y=-17] {Targeting}
\SetUpVertex[FillColor=black!20]
\Vertex[Math,L={Money\ Supply_{t^\ast-2}\ (L^{16}_{t^\ast-2})},x=4.5,y=-19] {MoneySupply1}
\SetUpVertex[FillColor=white]
%\Vertex[Math,L={Money\ Growth\ Target_{t^\ast}},x=7.5,y=-20] {GrowthTarget}

% Institutions and Quality
\SetUpVertex[FillColor=black!20]
\Vertex[Math,L={CBT_{t^\ast-2}\ (L^{13}_{t^\ast-2})},x=1,y=-12] {CBT}
\SetUpVertex[FillColor=red!20]
\Vertex[Math,L={CB\ Independence_{t^\ast-2}\ (A_{t^\ast-2})},x=0,y=-21] {CBI}
\SetUpVertex[FillColor=white]
\Vertex[Math,L={Time\ Preference_{t^\ast-2}},x=7.5,y=-21] {TimePref}
\Vertex[Math,L={Govern.\ Dec.{t^\ast-2}},x=-1.5,y=-18] {Government1}
%\Vertex[Math,L={Govern.\ Dec.{t^\ast-2}},x=-1.5,y=-5] {Government2}
\Vertex[Math,L={Int. Press.{t^\ast-2}},x=0,y=-11] {IntPress}
\SetUpVertex[FillColor=black!20]
\Vertex[Math,L={Pol.\ Instit._{t^\ast-2}\ (L^{14}_{t^\ast-2})},x=0,y=-9] {PolInstitution}
\Vertex[Math,L={Pol.\ Instab._{t^\ast-2}\ (L^{15}_{t^\ast-2})},x=1,y=-7] {Instability}
\Vertex[Math,L={GDP\ p.c._{t^\ast-2}\ (L^{12}_{t^\ast-2})},x=17.1,y=-18.75] {gdppc}
\SetUpVertex[FillColor=white]

% Money and Credit Market
\Vertex[Math,L={Money\ Demand_{t^\ast-2}},x=19,y=-26.75] {MoneyDemand}
\SetUpVertex[FillColor=black!20]
\Vertex[Math,L={Consumer\ Prices _{t^\ast-2}\ (L^{8}_{t^\ast-2})},x=12,y=-28] {PastPrices1}
\Vertex[Math,L={Money\ Supply_{t^\ast-2}\ (L^{16}_{t^\ast-2})},x=25,y=-27] {MoneySupply}
\SetUpVertex[FillColor=white]
\Vertex[Math,L={Nominal\ Exchange\ Rate_{t^\ast-2}},x=37,y=-21.5] {X-Rate}
%\Vertex[Math,L={Nominal\ X-Rate_{t^\ast-2}},x=31,y=-21.5] {X-RateT1}
\Vertex[Math,L={Nominal\ Interest\ Rate_{t^\ast-2}},x=25,y=-21.5] {InterestRate}
%\Vertex[Math,L={Nominal\ Interest\ Rate_{t^\ast-1}},x=37,y=-23.5] {InterestRateT1}
\Vertex[Math,L={Real\ Interest\ Rate_{t^\ast-2}},x=25,y=-20] {RealInterestRate}
\SetUpVertex[FillColor=black!20]
%\Vertex[Math,L={Inflation\ Expectations_{t^\ast-2}},x=31,y=-20] {PastExpectations}
\SetUpVertex[FillColor=white]
\Vertex[Math,L={MP\ Decision_{t^\ast-2}},x=14,y=-22] {MPDecision}
\SetUpVertex[FillColor=black!20]
\Vertex[Math,L={Capital\ Openness_{t^\ast-2}\ (L^{11}_{t^\ast-2})},x=1.5,y=-24.5] {CapitalOpenness}
\SetUpVertex[FillColor=white]
\Vertex[Math,L={Currency\ Competition_{t^\ast-2}},x=8.5,y=-23] {CurrencyCompetition}
\SetUpVertex[FillColor=black!20]
\Vertex[Math,L={Output_{t^\ast-2}\ (L^{5}_{t^\ast-2})},x=18.5,y=-28.5] {PastGDP}
\SetUpVertex[FillColor=white]
\Vertex[Math,L={Nominal\ Interest\ Rate_{t^\ast-3}},x=25.5,y=-28.5] {PastInterestRate}
%\Vertex[Math,L={Past\ Inflation_{t^\ast-2}},x=14.5,y=-25] {PastInflation}
\SetUpVertex[FillColor=black!20]
\Vertex[Math,L={Bank\ Loans_{t^\ast-2}\ (L^{18}_{t^\ast-2})},x=32.5,y=-23.5] {BankLoans}
\SetUpVertex[FillColor=white]
\Vertex[Math,L={Consumption_{t^\ast-1}},x=39.5,y=-23.5] {Consumption1}

%%%% Boxes %%%%

% Money and Credit Market
% \node[draw=red,dotted,thick,bigbox={Money \& Credit}, fit=(MoneyDemand)(MoneySupply)(InterestRate)(X-Rate)(MarketEfficiency)] {};
% Monetary Strategies
% \node[draw=red,dotted,thick,bigbox={Mon. Strategies}, fit=(ERA)(Targeting)] {};
% Public Finance
% \node[draw=red,dotted,thick,bigbox={Public Finance}, fit=(Fiscal Revenue)(PublicDebt)(Fiscal Spending)(BudgetBalance)(TASS1)(VAT1)] {};
% Fiscal Policy
% \node[draw=red,dotted,thick,bigbox={Public Finance}, fit=(VAT)(TASS)] {};
% Institutions
% \node[draw=red,dotted,thick,bigbox={Institutions}, fit=(CBT)(CBI)] {};
% \node[draw=red,dotted,thick,bigbox={Institutions}, fit=(Instability)(PolInstitution)(CBC)] {};
% Globalization
% \node[draw=red,dotted,thick,bigbox={Globalization}, fit=(TradeOpenness)(ShareNonTrade)(CapitalOpenness)(CurrencyCompetition)] {};
% Natural Ressources
% \node[draw=red,dotted,thick,bigbox={Natural Ressources}, fit=(Energy Price)] {};
% Demography
% \node[draw=red,dotted,thick,bigbox={Demography}, fit=(Ageing)] {};
% Inflation Memory
% \node[draw=red,dotted,thick,bigbox={Infl. Memory}, fit=(MovingAverage)] {};
% \node[draw=red,dotted,thick,bigbox={Infl. Memory}, fit=(MovingAverage1)] {};

%%%% Add. Back Door Paths: Priority 3 %%%%

\tikzset{EdgeStyle/.style={->,dashed}}

%\Vertex[Math,L={GDP\ p.c._{t^\ast-2}},x=35,y=-6] {A3}
%\Vertex[Math,L={Inst.\ Quality_{t^\ast-2}},x=42,y=-6] {A4}
%\Edge[label=$3.A$](A3)(A4)
%\Vertex[Math,L={Inst.\ Quality_{t^\ast-2}},x=35,y=-6.5] {A15}
%\Vertex[Math,L={Taxes\ and\ Social\ Securities_{t^\ast-1}},x=42,y=-6.5] {A16}
%\Edge[label=$3.B$](A15)(A16)
%\Vertex[Math,L={Inst.\ Quality_{t^\ast-2}},x=35,y=-7] {A17}
%\Vertex[Math,L={Market\ Power_{t^\ast-1}},x=42,y=-7] {A18}
%\Edge[label=$3.C$](A17)(A18)
%\Vertex[Math,L={Inst.\ Quality_{t^\ast-2}},x=35,y=-7.5] {A19}
%\Vertex[Math,L={Labour\ Unions_{t^\ast-1}},x=42,y=-7.5] {A20}
%\Edge[label=$3.D$](A19)(A20)

%\Edge[label=$3.B$](Targeting)(MoneySupply1)
%\Edge[label=$3.C$](PastDebt2)(MPDecision)

%%%% Edges %%%%

\tikzset{EdgeStyle/.style={->}}
\Edge[label=$1$](VAT)(Inflation)
\Edge[label=$2$](PC)(Inflation)
\Edge[label=$3$](Production Cost)(PC)
\Edge[label=$4$](Mark Up)(PC)
\Edge[label=$5$](Market Power)(Mark Up)
\Edge[label=$6$](GDP)(Mark Up)
\Edge[label=$7$](InterestRate)(Liquidity)
\Edge[label=$8$](Net Exports)(GDP)
\Edge[label=$9$](Fiscal Spending)(GDP)
\Edge[label=$10$](Consumption)(GDP)
\Edge[label=$11$](Investments)(GDP)
\Edge[label=$12$](Disposable Income)(Savings)
\Edge[label=$13$](Fiscal Spending)(RAD)
\Edge[label=$14$](Tobin)(Investments)
\Edge[label=$15$](Investments)(RAD)
 % ab jetzt sind gebogene Kanten, Winkel (0,360) in dem die Kanten rein und raus gehen.
%\tikzset{EdgeStyle/.style={->,relative=false,in=330,out=150}}
\Edge[label=$16$](GDP)(BC1)
 % Back to straight line
\tikzset{EdgeStyle/.style={->}}
\Edge[label=$17$](RADT2)(Machine Productivity1)
\Edge[label=$17$](RAD)(Machine Productivity1)
\Edge[label=$17$](RAD)(Machine ProductivityFuture)
 % ab jetzt sind gebogene Kanten, Winkel (0,360) in dem die Kanten rein und raus gehen.
\tikzset{EdgeStyle/.style={->,relative=false,in=250,out=175}}
\Edge[label=$18$](RAD)(LProductivity)
 % Back to straight line
\tikzset{EdgeStyle/.style={->}}
\Edge[label=$18$](RADT2)(LProductivity)
\Edge[label=$18$](RAD)(LProductivityFuture)
\Edge[label=$19$](Machine Productivity1)(BC1)

\Edge[label=$20$](Machine Productivity)(NLC)
\Edge[label=$21$](NLC)(Production Cost)
\Edge[label=$22$](Energy Price)(NLC)
\Edge[label=$23$](LC)(Production Cost)
\Edge[label=$24$](TASS)(LC)
\Edge[label=$25$](LProductivity)(EBP1)
\Edge[label=$26$](BC1)(EBP1)
\Edge[label=$27$](LUnion)(EBP1)
\Edge[label=$28$](EBP)(Nominal Wages)
\Edge[label=$29$](Nominal Wages)(LC)
 % ab jetzt sind gebogene Kanten, Winkel (0,360) in dem die Kanten rein und raus gehen.
\tikzset{EdgeStyle/.style={->,relative=false,in=340,out=0}}
\Edge[label=$30$](Targeting)(Inflation Expectation)
 % Back to straight line
\tikzset{EdgeStyle/.style={->}}
\Edge[label=$31$](Fiscal Revenue)(Fiscal Spending)
\Edge[label=$32$](CBC)(Inflation Expectation)
\Edge[label=$33$](Foreign Output)(Net Exports)
\Edge[label=$34$](Disposable Income)(Wealth)
\Edge[label=$35$](Last Nominal Wage)(Disposable Income)
\Edge[label=$36$](Fiscal Spending)(BudgetBalance)
\Edge[label=$37$](BudgetBalance)(PublicDebt)
\tikzset{EdgeStyle/.style={->,relative=false,in=220,out=20}}
\Edge[label=$38$](DebtManagement)(MPDecision)
 % Back to straight line
\tikzset{EdgeStyle/.style={->}}
\Edge[label=$39$](Last TSS)(Disposable Income)
\Edge[label=$40$](Past X-Rate)(Net Exports)
\tikzset{EdgeStyle/.style={->,relative=false,in=270,out=90}}
\Edge[label=$41$](Past Prices)(Past X-Rate)
\tikzset{EdgeStyle/.style={->}}
\Edge[label=$42$](Fiscal Revenue)(BudgetBalance)
\Edge[label=$43$](Savings)(Investments)
\Edge[label=$44$](MoneySupply)(InterestRate)
 % ab jetzt sind gebogene Kanten, Winkel (0,360) in dem die Kanten rein und raus gehen.
\tikzset{EdgeStyle/.style={->,relative=false,in=260,out=20}}
\Edge[label=$45$](MoneyDemand)(InterestRate)
 % Back to straight line
\tikzset{EdgeStyle/.style={->}}
\Edge[label=$46$](RealInterestRate)(Investments)
%\Edge[label=$47$](InterestRateT1)(X-Rate)
\Edge[label=$47$](InterestRate)(X-Rate)
\Edge[label=$48$](X-Rate)(Past X-Rate)
\Edge[label=$49$](TimePref)(MPDecision)
 % ab jetzt sind gebogene Kanten, Winkel (0,360) in dem die Kanten rein und raus gehen.
\tikzset{EdgeStyle/.style={->,relative=false,in=270,out=90}}
\Edge[label=$50$](Targeting)(CBC)
 % Back to straight line
 \tikzset{EdgeStyle/.style={->}}
\Edge[label=$51$](ERA)(CBC)
\Edge[label=$52$](MPDecision)(MoneySupply)
\Edge[label=$53$](PastDebt)(PublicDebt)
\tikzset{EdgeStyle/.style={->,relative=false,in=325,out=140}}
\Edge[label=$54$](MPDecision)(Targeting)
 % Back to straight line
 \tikzset{EdgeStyle/.style={->}}
\Edge[label=$55$](MPDecision)(ERA)
\Edge[label=$56$](MoneyDemand)(MPDecision)
\Edge[label=$57$](PastInterestRate)(MoneyDemand)
\Edge[label=$58$](PastGDP)(MoneyDemand)
 % ab jetzt sind gebogene Kanten, Winkel (0,360) in dem die Kanten rein und raus gehen.
\tikzset{EdgeStyle/.style={->,relative=false,in=343,out=75}}
\Edge[label=$59$](MPDecision)(Inflation Expectation)
 % Back to straight line
\tikzset{EdgeStyle/.style={->}}
\Edge[label=$60$](Wealth)(Consumption)
 % ab jetzt sind gebogene Kanten, Winkel (0,360) in dem die Kanten rein und raus gehen.
\tikzset{EdgeStyle/.style={->,relative=false,in=180,out=90}}
\Edge[label=$61$](Savings)(Wealth)
 % Back to straight line
\tikzset{EdgeStyle/.style={->}}
\Edge[label=$62$](RealInterestRate)(Assets)
\Edge[label=$63$](Assets)(Tobin)
 % ab jetzt sind gebogene Kanten, Winkel (0,360) in dem die Kanten rein und raus gehen.
%\tikzset{EdgeStyle/.style={->,relative=false,in=270,out=20}}
%\Edge[label=$64$](AssetsT1)(Savings)
%\Edge[label=$64$](Assets)(SavingsT2)
 % Back to straight line
\tikzset{EdgeStyle/.style={->}}
%\Edge[label=$65$](PastInflation)(MPDecision)
\Edge[label=$66$](MoneySupply)(BankLoans)
 % ab jetzt sind gebogene Kanten, Winkel (0,360) in dem die Kanten rein und raus gehen.
\tikzset{EdgeStyle/.style={->,relative=false,in=290,out=90}}
\Edge[label=$67$](BankLoans)(Investments)
 % Back to straight line
\tikzset{EdgeStyle/.style={->}}
\Edge[label=$68$](CBT)(CBC)
\Edge[label=$69$](CBI)(TimePref)
\Edge[label=$70$](TradeOpenness)(ShareNonTrade)
\Edge[label=$71$](ShareNonTrade)(MPDecision)
\Edge[label=$72$](MovingAverage1)(Inflation Expectation)
\tikzset{EdgeStyle/.style={->,relative=false,in=120,out=225}}
\Edge[label=$101$](MovingAverage1)(Government1)
\tikzset{EdgeStyle/.style={->}}
\tikzset{EdgeStyle/.style={->,relative=false,in=250,out=0}}
\Edge[label=$65$](MovingAverage)(MPDecision)
\tikzset{EdgeStyle/.style={->}}

%\Edge[label=$73$](PastInflation1)(Inflation Expectation)
\Edge[label=$74$](CBI)(CBC)
 % ab jetzt sind gebogene Kanten, Winkel (0,360) in dem die Kanten rein und raus gehen.
\tikzset{EdgeStyle/.style={->,relative=false,in=180,out=90}}
\Edge[label=$75$](Inflation Expectation)(Nominal Wages)
 % Back to straight line
\tikzset{EdgeStyle/.style={->}}
\Edge[label=$76$](Assets)(FirmValue)
\Edge[label=$77$](FirmValue)(MarketEfficiency)
\Edge[label=$78$](MarketEfficiency)(BankLoans)
\Edge[label=$79$](CapitalOpenness)(CurrencyCompetition)
\Edge[label=$80$](CurrencyCompetition)(MPDecision)
\Edge[label=$81$](PolInstitution)(CBC)
\Edge[label=$82$](Instability)(CBC)

\tikzset{EdgeStyle/.style={->,relative=false,in=110,out=190}}
\Edge[label=$59$](Instability)(Government1)
\tikzset{EdgeStyle/.style={->,relative=false,in=100,out=210}}
\Edge[label=$99$](PolInstitution)(Government1)
%\tikzset{EdgeStyle/.style={->,relative=false,in=260,out=90}}
\tikzset{EdgeStyle/.style={->}}
\Edge[label=$100$](IntPress)(PolInstitution)
\tikzset{EdgeStyle/.style={->,relative=false,in=260,out=90}}
\Edge[label=$86$](Government1)(CBT)
\tikzset{EdgeStyle/.style={->,relative=false,in=90,out=270}}
\Edge[label=$98$](Government1)(CBI)
\tikzset{EdgeStyle/.style={->}}

\Edge[label=$83$](gdppc)(MPDecision)
\Edge[label=$84$](Aging)(Savings)
\Edge[label=$85$](InterestRate)(RealInterestRate)
%\Edge[label=$86$](PastExpectations)(RealInterestRate)
\Edge[label=$87$](Liquidity)(MarketEfficiency)
\Edge[label=$88$](PastPrices1)(MoneyDemand)
\Edge[label=$89$](TASS1)(Fiscal Revenue)
\Edge[label=$90$](VAT1)(Fiscal Revenue)
\Edge[label=$91$](BudgetBalance)(Disposable Income)
 % ab jetzt sind gebogene Kanten, Winkel (0,360) in dem die Kanten rein und raus gehen.
\tikzset{EdgeStyle/.style={->,relative=false,in=340,out=90}}
\Edge[label=$92$](ERA)(Inflation Expectation)
 % Back to straight line
\tikzset{EdgeStyle/.style={->}}
\Edge[label=$93$](ERA)(MoneySupply1)
\Edge[label=$94$](Targeting)(MoneySupply1)
\Edge[label=$95$](PastDebt2)(MPDecision)
%----------------------------
\Edge[label=$96$](BankLoans)(Consumption1)
 % ab jetzt sind gebogene Kanten, Winkel (0,360) in dem die Kanten rein und raus gehen.
\tikzset{EdgeStyle/.style={->,relative=false,in=260,out=90}}
\Edge[label=$97$](RealInterestRate)(Savings)
 % Back to straight line
\tikzset{EdgeStyle/.style={->}}
 % ab jetzt sind gebogene Kanten, Winkel (0,360) in dem die Kanten rein und raus gehen.
\tikzset{EdgeStyle/.style={->,relative=false,in=240,out=20}}
\Edge[label=$64$](Assets)(Savings)
 % Back to straight line
\tikzset{EdgeStyle/.style={->}}
\tikzset{EdgeStyle/.style={->,relative=false,in=200,out=170}}
%\Edge[label=$99$](MovingAverage)(CBI)
\tikzset{EdgeStyle/.style={->}}
%\Edge[label=$33$](Inst)(LUnion)
%\Edge[label=$34$](Inst)(CBC)
%\Edge[label=$35$](Inst)(RAD)
\end{tikzpicture}
}
\caption{DAG containing the structural assumptions about the data generating process for a specific time point $t^{^\ast} = 2000,\ldots,2010$. The target quantity is $\psi_{j,k}$ and relates to $Y_{2010}$, which refers to \textit{$Consumer\ Prices_{t^\ast}$} colored in green. The intervention rules relate to CBI at time $t^{^\ast}-2$, colored in red. Measured covariates are grey, and unmeasured covariates are white. A justification of the DAG is given in Appendix \ref{sec:appendix_justification}.}
\label{fig:DAG}
\end{figure}
\end{landscape}

%% file: VertexReasoning.tex
%% Vertices Explanation
%\begin{table}
%\centering
\begin{longtable}{p{3cm}|p{12cm}|p{3cm}}
Node & Explanation & Emp. Approx. \\
\hline
\hline
\hline
  \rowcolor{Gray}
Consumer Prices & Price changes (\%) in the consumption basket of a representative household. & ConsumerPrices  \\
\hline
Consumption Tax & Value added tax on the net price of goods and services. & Unmeasured  \\
\hline
  \rowcolor{Gray}
Pricing by Companies & Firms set their product prices based on production costs and markups to maximize profit. & Unmeasured \\
\hline
Price Mark-Up & Surcharge on marginal cost. It depends on aggregate demand and market power. & Unmeasured   \\
\hline
  \rowcolor{Gray}
Production Cost & Convenient breakdown of unit costs into labor and
non-labor costs. It generally depends on the industry and countries' development. & Unmeasured   \\
\hline
Labor Costs & Direct wages, salaries, labor taxes, and social security contributions. & Unmeasured   \\
\hline
  \rowcolor{Gray}
Non-Labor Costs & Capital, land and intermediate inputs such as intermediate goods, primary commodities and energy. & Unmeasured   \\
\hline
Energy Prices & Weighted average of world market prices for energy resources such as oil, gas and coal measured in USD. & EnergyPrices \\
\hline
  \rowcolor{Gray}
Taxes and Social Security & Labor taxes and social security contributions. & Unmeasured \\
\hline
Market Power & Perfect competition forces firms to set marginal costs equal to prices. This corresponds to a lack of market power. By contrast, product differentiation suggests high market power. & Unmeasured  \\
\hline
  \rowcolor{Gray}
Output & Real Gross Domestic Product (GDP) measured in USD. In a small open economy, output consists of consumption, investments, government spending and net exports. &  Output \\
\hline
Consumption & Private consumption as a share of disposable household income. This is divided into two components: autonomous consumption and marginal propensity to consume. & Unmeasured  \\
\hline
  \rowcolor{Gray}
Disposable Income & Consumer income after transfers and taxes. & Unmeasured \\
\hline
Tobin's q & An economic measure that compares the market value of installed capital with the replacement cost of installed capital. A value greater than 1 leads to new investments. If the value is smaller than 1, purchasing existing capital is cheaper than investing in new capital. & Unmeasured \\
\hline
  \rowcolor{Gray}
Investments & Purchases of real estate by households and purchases of new capital goods (machines and plants) by firms. & Unmeasured  \\
\hline
Nominal Wages & Employees' salaries unrelated to the development of prices or indexation. & Unmeasured  \\
\hline
  \rowcolor{Gray}
Bargaining Power & Strength of bargaining position of employees in the
wage-setting process. & Unmeasured  \\
\hline
Labor Unions & Associations that represent the employed labor force in
setting wage levels, working conditions and worker rights. & Unmeasured \\
\hline
  \rowcolor{Gray}
Labor Productivity & The ratio of output (GDP) to the number of workers. & Unmeasured  \\
\hline
Output Gap & Fluctuations (\%) of current output (GDP) from its potential. & OutputGap  \\
\hline
  \rowcolor{Gray}
Technological Progress & A technological improvement resulting in higher machine productivity. & Unmeasured \\
\hline
Human and Public Capital & Expenses for discovering and developing new ideas and products. & Unmeasured \\
\hline
  \rowcolor{Gray}
Inflation Expectations & Expected consumer price level changes (\%) approximated by the backward-looking geometric mean of consumer price changes (\%)  over the past three years. & InflationExpectations  \\
%\hline
%Inflation Memory & Past consumer price level changes influence inflation expectations, in particular periods of either deflation or hyperinflation. & past\_infl &  L \\
\hline
Savings & The sum of accumulated private (and public) savings. Savings can be negative.  & Unmeasured \\
\hline
  \rowcolor{Gray}
Foreign Output & World output (GDP) depending on foreign consumption, investment and fiscal spending. Measured in USD as the sum of outputs for every country in the sample minus the output of the particular country. & ForeignOutput  \\
\hline
Net Exports & Defined as exports minus the value of imports. & Unmeasured  \\
\hline
  \rowcolor{Gray}
Real Exchange Rate & Determined by the nominal exchange rate and the domestic and foreign price levels. & Unmeasured  \\
\hline
Nominal Exchange Rate & Domestic currency in terms of foreign currency. & Unmeasured  \\
\hline
  \rowcolor{Gray}
Fiscal Spending & The sum of all government expenditures (on education, consumption, investments, etc.). & Unmeasured \\
\hline
Fiscal Revenue & The sum of fiscal earnings (mainly taxes). & Unmeasured \\
\hline
  \rowcolor{Gray}
Primary Balance & Primary surplus/deficit: Government revenues minus government spending excluding interest payments on outstanding debt in percent of GDP. & PrimaryBalance  \\
\hline
Public Debt & If the government runs a primary deficit in a given year, debt increases. The increase in debt is exacerbated by interest payments on existing debt. Measured in in percent of GDP. & PublicDebt \\
\hline
  \rowcolor{Gray}
Debt Management & Decisions of a government on debt structure, potentially resulting in different currency, price and interest-rate indexation composition as well as different maturities of newly issued and outstanding debt. & Unmeasured \\
\hline
Money Demand & Demand for money, defined as currency plus deposit accounts, determined by GDP and the level of interest rates on bonds. & Unmeasured  \\
\hline
  \rowcolor{Gray}
Money Supply & Different monetary aggregates (M0-M3) are available. For this analysis M2 was used. Measured average annual growth rate (\%) in money and quasi money. & MoneySupply \\
\hline
Nominal Interest Rate & The level of the interest rate is determined by the intersection of money supply and money demand. & Unmeasured  \\
\hline
\rowcolor{Gray}
Targeting Regime & Monetary policy strategy introduced in the 1990s
intended to stabilize inflation at a pre-announced point target or target range. & Unmeasured \\
\hline
Exchange-Rate Regime & Monetary policy strategy intended to stabilize inflation at a level commensurate with that of a strong currency. By pegging the currency to an anchor country's currency, its monetary policy and, hence, inflation is imported. Deviations from the target exchange rate are corrected by purchases and sales of the pegged currency. & Unmeasured \\
\hline
  \rowcolor{Gray}
Capital Openness & Index measuring a country's degree of capital account openness. & CapitalOpenness \\
\hline
AS \& MH & Adverse selection and moral hazard due to information asymmetries in credit markets. & Unmeasured \\
\hline
  \rowcolor{Gray}
Firms' net worth & A firm's total assets minus its total liabilities yields its equity. & Unmeasured \\
\hline
Firms' liquidity & Firms' liquidity is directly linked to their cashflow. Cash is the most liquid asset and is used to meet short-term liabilities. & Unmeasured \\
\hline
\rowcolor{Gray}
Age structure & Demographic indicator that captures the share of the total population older than 65 years. & AgeStructure \\
\hline
Trade openness & The sum of imports and exports is set in relation to a country's output (\%). It is a proxy for globalization. & TradeOpenness  \\
%\hline
%Int. Competition & In open economies firms compete with foreign companies for the favor of consumers. Under more trade openness and little trade frictions this accelerates global competition. & Unmeasured & U \\
\hline
  \rowcolor{Gray}
Asset Prices & Prices of assets in which households, firms, or governments are able to hold wealth, such as stocks, bonds, bank deposits, cash or real estate. & Unmeasured \\
\hline
Real Interest Rate & The difference between the nominal interest rate and the expected rate of inflation. & Unmeasured  \\
\hline
  \rowcolor{Gray}
Currency Competition & Governments and central banks are forced to implement disciplined policies since they compete with foreign currencies for capital. The primary mechanism through which greater openness to foreign capital might lead to lower inflation arises presumably from its disciplining effect on monetary policy. & Unmeasured  \\
\hline
CBT & Central banks publicly announce their forecasts, policy decisions and assessments of the economy. A central bank's transparency is strongly related to its accountability and its credibility. Its measured by means of a numeric index ranging from 0 (least transparent) to 15 (most transparent). & CBTransparency \\
\hline
  \rowcolor{Gray}
CB Independence & Independence of a central bank from governmental
bodies. Measured via de jure indices (e.g., statutes); see the main text for detailed explanations. Dichotomized. & CBIndependence  \\
\hline
CB Credibility & A central bank that does what it has announced publicly is considered to be credible. This is reflected in inflation expectations that are low and stable. & Unmeasured  \\
%\hline
%FP Decision & A decision made by the government concerning the use of the governments' budget to affect the volume of national spending, or more generally to provide public goods and services, as well as to redistribute income. & Unmeasured & L  \\
\hline
  \rowcolor{Gray}
Pol. Instab. & The percentage of veto players dropping from the government in any given year. In presidential systems, veto players are defined as the president and the largest party in the legislature. In parliamentary systems, the veto players are defined as the prime minister and the three largest government parties. & PolInstability \\
\hline
Pol. Instit. & A variety of dimensions: the stability orientation of society, the age of a country’s political parties, average government duration, democratization and economic liberalization, the degree of civil liberties, the presence of a higher chamber and federalism as well as a legislative process characterized by extensive checks and balances. Measured as a ordered categorical variable with seven categories where "7" represents the lowest degree of civil liberties and "1" the highest. & PolInstitution \\
\hline
  \rowcolor{Gray}
Time Preference & Time horizon envisaged by policymakers within which they want to achieve a certain macroeconomic outcome. It may vary from a short (high time preference) to a middle- to long-term perspective (low time preference). & Unmeasured  \\
\hline
Share of Non-Tradables & Distinction between tradeable and non-tradeable goods. Non-tradability means that a good is produced and consumed in the same economy (e.g. haircuts). & Unmeasured   \\
\hline
  \rowcolor{Gray}
GDP p.c. & GDP is the sum of all finished goods and services that are produced in a year. The p.c. term divides this value by the number of citizens. GDP p.c. is a proxy for economic wealth and living standards and is measured in USD. & GDPpc  \\
\hline
Bank Loans & Commercial banks create money when they offer loans depending on the availability of central bank reserves at their disposal. Measured as domestic credit to private sector (\% of GDP). & BankLoans  \\
\hline
\rowcolor{Gray}
Past Inflation & Consumer price changes (\%) during the past 7 years. "PastInflation" takes the median over those years. & PastInflation  \\
\hline
MP Decision & Monetary policy makers' (i.e., central bankers') decisions are contractionary, neutral or expansionary. & Unmeasured   \\
\hline
\rowcolor{Gray}
Wealth & Household wealth is accumulated savings over previous periods
(it can be negative in the event of net debt) and disposable income in the current period. & Unmeasured \\
\hline
Int. Pressure & International institutions or foreign aids can provide an important incentive to reform the central bank. For example the IMF regularly demands countries to end monetary financing of public debt, remove central bank governors and board members, and sometimes even pushes for full-fledged central bank reform. Further pressure may be exerted by the (regional) peer CBI community and international investors. & Unmeasured \\
\hline
\rowcolor{Gray}
Govern. Dec. & Captures all decisions a government takes but which are limited to CBI. & Unmeasured

\label{tab:VerticesDescription}
\end{longtable}

%% file: EdgeReasoning.tex
%% Edge Explanation
%\begin{table}
%\centering
\begin{longtable}{p{1cm}|p{14cm}|p{3cm}}
Arrow  & Causality Assumption & Source \\
\hline
\hline
\hline
  \rowcolor{Gray}
1 & Consumer prices can change after changes in consumption taxes (e.g., VAT). & \citet{gelardiTax} \\
\hline
2 & Consumer prices are set individually by retailers and companies. & \citet[p. 290]{burda2010macro}  \\
\hline
  \rowcolor{Gray}
3 & Production costs generally dominate the price-setting
process. Profit margins strongly depend on the industry in question. & \citet[p. 291]{burda2010macro}  \\
\hline
4 & Channels the aggregate demand side of the price-setting process. In a small open economy, demand shocks to goods and services affect the price level. & \citet[p. 312]{burda2010macro}  \\
\hline
  \rowcolor{Gray}
5 & Higher product differentiation leads to higher market power and higher markups in a profit-maximizing environment. & \citet[p. 291]{burda2010macro}  \\
\hline
6 & Changes in aggregate demand in the goods market enable firms to set higher prices. & \citet{bloch2001pricing} \\
\hline
  \rowcolor{Gray}
7 & Expansionary monetary policy, which lowers nominal interest rates, also causes an improvement in firms' balance sheets because it raises their cash flow. The rise in cash flow increases firms' (or households') liquidity. & \citet[p. 544 f.]{mishkin2013economics} \\
\hline
8 & In a small open economy, domestic demand for goods, and thus output, is also affected by net exports. & \citet[p. 125]{blanchard2010macro}  \\
\hline
  \rowcolor{Gray}
9 & Fiscal spending describes the decision of the government to spend money. It affects output (GDP). & \citet[p. 45]{blanchard2010macro}  \\
\hline
10 & Private consumption also affects output. & \citet[p. 44]{blanchard2010macro}  \\
\hline
  \rowcolor{Gray}
11 & Investments are another factor affecting output. & \citet[p. 44]{blanchard2010macro}  \\
\hline
12 & The share of disposable income that is not consumed in this period is saved based on the marginal propensity to save. & \citet[p. 52]{blanchard2010macro}  \\
\hline
  \rowcolor{Gray}
13 &  Governments undertake investments in human capital (e.g., education) or public capital (e.g., infrastructure) to bolster long-run economic growth. & \citet[pp. 85 ff.]{burda2010macro} \\
\hline
14 & A Tobin's q not equal to 1 gives incentives to invest or divest in capital and therefore affects aggregate investment. & \citet[p. 195]{burda2010macro}  \\
\hline
  \rowcolor{Gray}
15 & Similar to arrow 13, companies and other non-governmental agents affect human capital. & \citet[pp. 85 ff.]{burda2010macro}  \\
\hline
16 & The current value of GDP may deviate from its potential. & \citet[p. 11]{burda2010macro}  \\
\hline
  \rowcolor{Gray}
17 & Investments in human capital have a positive impact on innovation and economic development. & \citet{diebolt2019long} \\
\hline
18 & Training and education generally lead to high-skilled workers, and in turn, to high productivity of the labor force. & \citet[pp. 85 f.]{burda2010macro} \\
\hline
  \rowcolor{Gray}
19 & Potential output growth is mainly determined by technological progress. & \citet[p. 71]{burda2010macro} \\
\hline
20 & Technological progress indicates higher productivity, and higher productivity can again be expressed as obtaining the same output with fewer inputs (here, lower non-labor costs and higher profits) & \citet[p. 71]{burda2010macro} \\
\hline
  \rowcolor{Gray}
21 & The first (second cf. 23) component that determines the production costs are non-labor costs. & \citet[p. 291]{burda2010macro}  \\
\hline
22 & Changes in energy prices are transmitted through supply shocks
and affect the non-labor costs of production. &
\citet[p. 297]{burda2010macro}  \\ 
\hline
  \rowcolor{Gray}
23 & The second component that determines production costs are labor costs. & \citet[p. 291]{burda2010macro}  \\
\hline
24 & Gross hourly labor costs also include vacation, social security contributions and other benefits paid by employers to the benefit of workers. & \citet[p. 291]{burda2010macro}  \\
\hline
  \rowcolor{Gray}
25 & Higher skills increase workers' bargaining power in the wage-setting process. & \citet{cahuc2006wage}  \\
\hline
26 & During boom periods, rising employment generally improves the bargaining position of workers. & \citet[p. 294]{burda2010macro}  \\
\hline
  \rowcolor{Gray}
27 & Labor unions generally improve the bargaining position of workers. & \citet[p. 121]{burda2010macro}  \\
\hline
28 & A better bargaining position leads to higher wage markup. & \citet[p. 294]{burda2010macro}  \\
\hline
  \rowcolor{Gray}
29 & Nominal wages translate directly into labor costs. & \citet[p. 292]{burda2010macro}  \\
\hline
30 & Inflation expectations are built on publicly announced inflation targets. & \citet{gurkaynak2010does}  \\
\hline
  \rowcolor{Gray}
31 & Fiscal revenue increases the government's capacity to spend. & \citet[p. 136]{walsh2010monetary}  \\
\hline
32 & A credible central bank commitment to low and stable inflation anchors long-run inflation expectations. & \citet{BernankeTargeting}  \\
\hline
  \rowcolor{Gray}
33 & Net exports depend positively on foreign output. & \citet[p. 125]{blanchard2010macro}  \\
\hline
34 & Wealth depends on disposable income. & \citet[p. 136]{heijdra2002foundations}  \\
\hline
  \rowcolor{Gray}
35 & Disposable income is also determined by wages. & \citet[p. 43]{blanchard2010macro}  \\
\hline
36 & Fiscal spending affects the primary balance. & \citet[p. 439]{blanchard2010macro}  \\
\hline
  \rowcolor{Gray}
37 & Current primary deficits are financed by new debt. & \citet[p. 12]{burnside2005fiscal}  \\
\hline
38 & Missale and Blanchard (1994) introduced what they called “effective maturity”. Effective maturity measures the sensitivity of debt to unexpected inflation. The lower it is, the lower the impact of surprise inflation on the value of the debt and the lower the incentive to inflate. & \citet{EffectiveMaturity}  \\
\hline
  \rowcolor{Gray}
39 & Disposable income decreases following an increase in taxes and social security contributions s and increases after a reduction in taxes and social security contributions. & \citet{OECD_Tax}\\
\hline
40 & Exports depend negatively on the real exchange rate.  & \citet[p. 125]{blanchard2010macro}  \\
\hline
  \rowcolor{Gray}
41 & The real exchange rate is partly determined by the domestic price level. & \citet[p. 112]{blanchard2010macro}  \\
\hline
42 & Fiscal revenue affects the primary balance. & \citet[p. 439]{blanchard2010macro}  \\
\hline
  \rowcolor{Gray}
43 & Investments are proportional to output. Higher output implies higher savings and, thus, higher investments. & \citet[p. 248]{blanchard2010macro}  \\
\hline
44 & The interest rate is determined by the equilibrium condition that the supply of money be equal to the demand for money. & \citet[p. 77]{blanchard2010macro}  \\
\hline
  \rowcolor{Gray}
45 & The interest rate is determined by the equilibrium condition that the supply of money be equal to the demand for money. & \citet[p. 77]{blanchard2010macro}  \\
\hline
46 & "An important feature of the interest-rate transmission mechanism is its emphasis on the real (rather than the nominal) interest rate as the rate that affects consumer and business decisions. (\dots) lower real interest rates then lead to rises in business fixed investment, residential housing investment, inventory investment and consumer durable expenditure, (\dots)". & \citet[p. 537]{mishkin2013economics}  \\
\hline
\rowcolor{Gray}
47 & Investors face a choice between domestic and foreign assets and choose the investment with the highest expected return. & \citet[p. 119]{blanchard2010macro}   \\
\hline
48 & The nominal exchange rate is fundamental to the determination of the real exchange rate. & \citet[p. 112]{blanchard2010macro}   \\
\hline
\rowcolor{Gray}
49 & The degree of central bank independence plays a meaningful role only if the central bank places a different emphasis on alternative policy objectives than the government. The literature points to two main differences. One relates to possible differences between the rate of time preference of political authorities and that of central banks. For various reasons, central banks are often more conservative and take a longer view of the policy process than do politicians. The other difference concerns the subjective weights in the objective function of the central bank and that of the government. It is often assumed that central bankers are more concerned about inflation than about policy goals such as the achievement of high employment levels and adequate government revenues. &  \citet[p. 7]{eijffinger1996political}   \\
\hline
50 & Central banks publicly communicate their inflation target or range. One benefit of IT adoption is "a well-known and credible inflation target helps to anchor the private sector's long-run inflation expectations". & \citet[p. 1248]{svensson2010inflation}  \\
\hline
  \rowcolor{Gray}
51 & Central banks publicly communicate when they peg their currency, which affects their credibility. & \citet[pp. 492 f.]{burda2010macro}  \\
\hline
52 & Central bank operations affect the money supply. & \citet[pp. 301 ff.]{mishkin2013economics}   \\
\hline
\rowcolor{Gray}
53 & The government's flow budget constraint means that current government debt is dependent on past debt (and other items). & \citet[p. 36]{burnside2005fiscal} \\
\hline
54 & Given the breakdown of the relationship between monetary aggregates and goal variables such as inflation, many countries have recently adopted inflation targeting as their monetary policy regime. & \citet[pp. 590 f.]{mishkin1999international} \\
\hline
  \rowcolor{Gray}
55 & Targeting the exchange rate is a monetary policy regime with a
long history of adoption by central banks. & \citet[p. 581]{mishkin1999international}  \\
\hline
56 & Monetary authorities react to changes in the demand for money. &  \citet[pp. 216-7]{burda2010macro}  \\
\hline
  \rowcolor{Gray}
57 & Demand for the monetary base M0 (money produced by the central bank) depends negatively on the nominal interest rate. &  \citet[p. 217]{burda2010macro}  \\
\hline
58 & Demand for the monetary base M0 depends positively on nominal GDP. &  \citet[p. 217]{burda2010macro}  \\
\hline
  \rowcolor{Gray}
59 & Masciandaro and Romelli (2019) find that more stable governments are associated with higher levels of CBI. Cukierman and Webb (1995) show that CBI is lower in less stable political systems &  \citet[p. 217]{cukierman.1995, masciandaroRomelli.2019}  \\
\hline
% 59 & Central banks have increasingly issued information concerning statistics, forecasts and assessments of the economy to the public in the last two decades. & See central banks' websites for examples. \\
% \hline
60 & "When stock prices rise, the value of financial wealth increases, thereby increasing the lifetime resources of consumers, and consumption should rise." & \citet[p. 542]{mishkin2013economics}\\
\hline
61 & Savings lead to higher wealth. & \citet{cooper2016wealth} \\
\hline
  \rowcolor{Gray}
62 & "\dots, when monetary policy is expansionary, the public finds that it has more money than it wants and so gets rid of it through spending. One place the public spends is in the stock market, increasing the demand for stocks and consequently raising their prices." & \citet[p. 542]{mishkin2013economics} \\
\hline
63 & Tobin defines q as the market value of firms divided by the replacement cost of capital. & \citet[p. 540]{mishkin2013economics}  \\
\hline
  \rowcolor{Gray}
64 & Asset returns have a significant effect on household savings. & \citet{disney2010house}  \\
\hline
65 & Central banks' main objective is stable and low inflation. When inflation exceeds, or is expected to exceed, a certain level, a reaction by the central bank follows. & \citet{taylor1993discretion}  \\
\hline
  \rowcolor{Gray}
66 & "\dots the bank lending channel of monetary transmission operates as follows: expansionary monetary policy, which increases bank reserves and bank deposits, increases the quantity of bank loans available." & \citet[pp. 542 f.]{mishkin2013economics}  \\
\hline
67 & "Because many borrowers are dependent on bank loans to finance their activities, this increase in loans will cause investment (and possibly consumer) spending to rise \dots." & \citet[pp. 542 f.]{mishkin2013economics}  \\
\hline
    \rowcolor{Gray}
68 & Central bank transparency is multidimensional, covering political transparency (openness about policy objectives), economic transparency (openness about data, models, and forecasts), procedural transparency (openness about the way decisions are made, achieved mainly through the release of minutes and votes), policy transparency (openness about the policy implications, achieved through prompt announcement and explanation of decisions), and operational transparency (openness about the implementation of those decisions). Transparency is a means of enhancing the credibility of central bank commitments. & \citet{dincer2014central}  \\
\hline
69 & The most prominent argument for central bank independence is based on the time-inconsistency problem. It arises when the best plan made in the present for some future period is no longer optimal when that period actually starts. Implicitly, CBI reduces the time preference of monetary policy makers.& \citet[p. 5]{eijffinger1996political}  \\
\hline
  \rowcolor{Gray}
70 & When a country becomes more open in economic terms, the nontraded sector becomes less important than the traded goods sector. & \citet{LaneOpenness}  \\
\hline
71 & The more important the traded good sector is, the less that
monetary authorities stand to gain from surprise inflation because a
monetary expansion in an open economy will be accompanied by a real
depreciation of the currency, raising costs for households and
businesses. The larger the share of imported goods is, the greater the increase in inflation. & \citet{LaneOpenness} and \citet{romer1993openness}  \\
\hline
  \rowcolor{Gray}
72 & \textit{Past Inflation} can be considered as a summary statistic of past consumer price movements. & By definition.  \\
\hline
73 & "The hybrid Phillips curve is an example of how models used in the policy arena seek to overcome unsatisfactory features of both the adaptive expectations Phillips curve (it is empirically successful, but is subject to the Lucas critique; lacks micro-foundations and rational expectations; and lacks a channel for credibility to affect inflation) and the NKPC (which is forward looking and therefore not subject to the Lucas critique; has micro-foundations and rational expectations with a role for credibility, but counterfactual empirical predictions). The hybrid Phillips includes forward-looking inflation expectations but acknowledges that inflation appears to be persistent or inertial, i.e. that it depends on lagged values of itself....The hybrid Phillips curve can be rationalized by the assumption that some proportion of firms use a backward-looking rule of thumb to set their inflation expectations while the remainder use forward-looking expectations." & \citet[p. 610]{SoskiceCarlin} \\
\hline
  \rowcolor{Gray}
74 & One way for a central bank to establish credibility is by increasing its independence. & \citet{blinder2000central}  \\
\hline
75 & Employees want to protect themselves from a loss in purchasing power, so they embed their inflation expectations into their nominal wages. & \citet[p. 293]{burda2010macro} \\
\hline
  \rowcolor{Gray}
76 & "Expansionary monetary policy, which causes a rise in stock prices along the lines described earlier, raises the net worth of firms \dots". & \citet[p. 544]{mishkin2013economics} \\
\hline
77 & "The lower the net worth of business firms, the more severe the adverse selection and moral hazard problems in lending to these firms. Lower net worth means that lenders in effect have less collateral for their loans, so their potential losses from adverse selection are higher.". & \citet[p. 544]{mishkin2013economics} \\
\hline
  \rowcolor{Gray}
78 & "The lower net worth of businesses also increases the moral hazard problem because it means that owners have a lower equity stake in their firms, giving them more incentive to engage in risky investment projects. Because taking on riskier investment project makes it more likely that lenders will not be paid back, a decrease in businesses’ net worth leads to a decrease in lending and hence in investment spending.". & \citet[p. 544]{mishkin2013economics} \\
\hline
79 & In a more integrated world, competition between currencies is even more present since countries want to attract foreign investments, and this race is exacerbated in a financially integrated world. & \citet{captialOpenness} \\
\hline
  \rowcolor{Gray}
80 & The primary mechanism through which greater openness to foreign capital might lead to lower inflation is presumably some sort of disciplining effect on monetary policy.& \citet{captialOpenness} \\
\hline
81 & The quality of political institutions might directly influence the relationship between CBI and inflation. The effectiveness of CBI in strengthening credibility and enhancing inflation performance is increased by the presence of multiple political veto players or if checks and balances are sufficiently strong. & \citet{keefer_stasavage_2003, Hayo2008} \\
\hline
  \rowcolor{Gray}
82 & Political instability can have a number of possible effects. The
most commonly discussed of these is that more instability makes it difficult for policy makers to commit to low inflation. & \citet[p. 10]{CampilloMiron}  \\
\hline
83 & Income per capita captures several possible effects. A higher
level of income per capita is likely to be accompanied by a more
sophisticated tax system and a more developed financial system, both
of which imply a lower optimal inflation tax and thus a negative relation with inflation. On the other hand, high-income countries might be better at innovating technologies for reducing the costs of inflation, so their inflation aversion might be lower. & \citet[p. 11]{CampilloMiron} \\
\hline
  \rowcolor{Gray}
84 & The life-cycle theory suggests that individuals plan their
consumption and savings behavior over their life-cycle and smooth out
their consumption over their lifetimes. Aggregate demand and supply shift because certain age groups and their particular economic behavior gain in relative importance to the rest of the population. Hence, changes in the demographic structure can exert potentially large effects on total savings. & \citet[p. 5]{bobeica2017demographics} \\
\hline
85 & For given prices, nominal and real interest rates are directly linked through the Fisher equation. & \citet[p. 524]{burda2010macro}  \\
\hline
  \rowcolor{Gray}
86 & The government introduces or changes the law of the central bank that affects its obligations to provide more transparency.
 & By definition\\
\hline
87 & "Another balance sheet channel operates by affecting cash flow, the difference between cash receipts and cash expenditures. The rise in cash flow increases the liquidity of the firm (or household) and thus makes it easier for lenders to know whether the firm (or household) will be able to pay its bills. The result is that adverse selection and moral hazard problems become less severe, \dots".  & \citet[p. 544 f.]{mishkin2013economics} \\
\hline
  \rowcolor{Gray}
88 & Money demand depends on nominal output, so the price level becomes relevant. & \citet[p. 217]{burda2010macro} \\
\hline
89 \& 90 & The government also collects its revenue through tax payments. & \citet[p. 136]{walsh2010monetary}  \\
\hline
  \rowcolor{Gray}
91 & If the government runs a budget deficit by holding spending constant and reducing tax revenue, households’ current disposable income, and perhaps their lifetime wealth, increases. & \citet{Debt}  \\
\hline
92 & If the exchange-rate target is credible, it anchors inflation expectations to the inflation rate in the anchor country to whose currency it is pegged. & \citet[p. 581]{mishkin1999international}  \\
\hline
  \rowcolor{Gray}
93 & Pegging the exchange rate to a foreign anchor forces the country to adopt the foreign interest rate policy, which affects broad domestic money supply. & \citet{mishkin1999international}  \\
\hline
94 & In an inflation (forecast) targeting framework, the central bank
changes its short-term interest rate if the inflation forecast exceeds
or falls short of the inflation target, until the inflation forecast
equals the target. In a related version of the inflation targeting
strategy, the central bank may deem it appropriate to adjust its monetary policy if the inflation forecast indicates a deviation from target (or its range). In either case, the money supply will be affected. & \citet{svensson1997inflation} and \citet{peytrignet2007money}  \\
\hline
  \rowcolor{Gray}
95 & A government that issues nominal debt has an incentive to promise low inflation ex ante to lower nominal interest payments and then reduce the ex post value of the debt through unexpected inflation. This incentive is stronger the larger the public debt is. & \citet{Kwon2009}  \\
\hline
96 & Credit growth is a more important determinant of consumption than income growth. & \cite{ConsumptionCredit}  \\
\hline
  \rowcolor{Gray}
97 & Capital and savings are usually valued by discounting. The amount
of the discount depends primarily on the real interest rate. &  \citet[p. 161]{burda2010macro}    \\
% \hline
% 98 & Savings are typically invested in equities or shares that are valued at stock exchanges or other equity markets. &  \citet[p. 346-9]{burda2010macro}    \\
\hline
98 & The government legislates on the degrees of independence to be granted to the central bank. & By definition\\
\hline
  \rowcolor{Gray}
99 & Governments in federally organized countries with good checks and balances grant their monetary institutions greater autonomy (De Haan and van't Hag (1995), Moser (1999), Keefer and Stasavage (1999) or Farvaque (2002)). Reforms are also more likely following elections that lead to a political consolidation or to changes in the political orientation of the government (Romelli, 2018). Masciandaro and Romelli (2019) find evidence that countries in common-law jurisdictions have a lower degree of independence. &  \citet{deHaan.95, Moser.1999, farvaque.2002, keefer_stasavage_2003, romelli.2018, masciandaroRomelli.2019}   \\
\hline
100 & Binding agreements with international lenders like the IMF or the World Bank often require countries to commit to a particular set of policies (Blejer et al., 2002; Gutierrez, 2003; Polillo and Guillén, 2005; Rodrik and Bank, 2006; Romelli, 2018; Kern, Reinsberg and Rau-Göhring, 2019; Reinsberg et al. (2020)). Another type of external pressure can come from regional clustering, which is often found to be cohesive of certain types of reform processes such as democratisations and economic liberalisations (Simmons and Elkins, 2004; Elhorst et al., 2013; Giuliano et al., 2013; Acemoglu et al., 2019). & \citet{blejer2002inflation, gutierrez2003inflation, simmons2004globalization, polillo2005globalization, rodrik2006goodbye, elhorst2013impact, giuliano2013democracy, romelli.2018, kern2019imf, acemoglu2019democracy, reinsberg2020bad}
\\
\hline
  \rowcolor{Gray}
101 & Masciandaro and Romelli (2019) find that countries experiencing high inflation in the past have a higher inflation aversion which constrains the government to assign a higher degree of CBI in the following periods. Similarly, Crowe and Meade (2008) show that over the period 1990–2003, greater changes in CBI have occurred in countries with higher inflation. According to Wachtel and Blejer (2020) the arguments in favor of an independent central bank began to crystallize in the 1980s after a decade or more of traumatic inflationary experience that put a spotlight on central bank policymaking and its failures. & \citet{crowe2007evolution, masciandaroRomelli.2019, wachtel.2020}

\label{tab:Edge Description}
\end{longtable}